\documentclass[%
twocolumn,
superscriptaddress,
 amsmath,amssymb,
pra,
floatfix,
]{revtex4-2}

\usepackage[english]{babel}
\usepackage[utf8]{inputenc}
\usepackage{dcolumn}% Align table columns on decimal point
\usepackage{bm}% bold math
\usepackage{amsthm}
\usepackage{amssymb}
\usepackage{mathtools}
\usepackage{physics}
\usepackage{xcolor}
\usepackage{graphicx}
\usepackage{adjustbox}
\usepackage{placeins}
\usepackage[T1]{fontenc}
\usepackage{lipsum}
\usepackage{csquotes}
\usepackage{comment}
\usepackage{xr}
\usepackage{makecell}
\usepackage{ mathrsfs }
\usepackage{url}
\usepackage{capt-of}

\newcommand{\dnv}{\ensuremath{d_{\mathrm{NV}}}}
\newcommand{\numwunit}[2]{\ensuremath{#1 \, \mathrm{#2}}}
\newcommand{\thetanv}{\ensuremath{\vartheta_{\mathrm{NV}}}}
\newcommand{\phinv}{\ensuremath{\varphi_{\mathrm{NV}}}}
\newcommand{\unv}{\ensuremath{\hat{u}_{\mathrm{NV}}}}
\newcommand{\unvphi}{\ensuremath{\hat{u}_{\mathrm{NV},\varphi}}}

\newcommand{\timeder}[1]{\dfrac{\mathrm{d} #1}{\mathrm{d} t}}

\makeatletter
\newcommand*{\addFileDependency}[1]{% argument=file name and extension
  \typeout{(#1)}
  \@addtofilelist{#1}
  \IfFileExists{#1}{}{\typeout{No file #1.}}
}
\makeatother

\begin{document}

\title{Multi-angle reconstruction of domain morphology with all-optical diamond magnetometry}

\author{Lucio Stefan}
\thanks{Contributed equally to this work}
\email{lucio.stefan@nbi.ku.dk}
\affiliation{Cavendish Laboratory, University of Cambridge, J. J. Thomson Avenue, Cambridge, CB3 0HE, UK}
\affiliation{The Faraday Institution, Quad One, Becquerel Avenue, Harwell Campus, Didcot, OX11 0RA, UK}

\author{Anthony K. C. Tan}
\thanks{Contributed equally to this work}
\affiliation{Cavendish Laboratory, University of Cambridge, J. J. Thomson Avenue, Cambridge, CB3 0HE, UK}

\author{Baptiste Vindolet}
\thanks{Contributed equally to this work}
\affiliation{Universit\'e Paris-Saclay, CNRS, ENS Paris-Saclay, CentraleSup\'elec, LuMIn, 91190, Gif-sur-Yvette, France}

\author{Michael H\"ogen}
\affiliation{Cavendish Laboratory, University of Cambridge, J. J. Thomson Avenue, Cambridge, CB3 0HE, UK}

\author{Dickson Thian}
\affiliation{Institute of Materials Research and Engineering, Agency for Science, Technology and Research (A*STAR), 138634 Singapore}

\author{Hang Khume Tan}
\affiliation{Institute of Materials Research and Engineering, Agency for Science, Technology and Research (A*STAR), 138634 Singapore}

\author{Lo\"ic Rondin}
\affiliation{Universit\'e Paris-Saclay, CNRS, ENS Paris-Saclay, CentraleSup\'elec, LuMIn, 91190, Gif-sur-Yvette, France}

\author{Helena S. Knowles}
\affiliation{Cavendish Laboratory, University of Cambridge, J. J. Thomson Avenue, Cambridge, CB3 0HE, UK}

\author{Jean-Fran\c{c}ois Roch}
\affiliation{Universit\'e Paris-Saclay, CNRS, ENS Paris-Saclay, CentraleSup\'elec, LuMIn, 91190, Gif-sur-Yvette, France}

\author{Anjan Soumyanarayanan}
\affiliation{Institute of Materials Research and Engineering, Agency for Science, Technology and Research (A*STAR), 138634 Singapore}
\affiliation{Physics Department, National University of Singapore (NUS), 117551 Singapore}

\author{Mete Atat\"ure}
\email{ma424@cam.ac.uk}
\affiliation{Cavendish Laboratory, University of Cambridge, J. J. Thomson Avenue, Cambridge, CB3 0HE, UK}

%%%%%%%%%%%%%%%%%%%%%%%%%%%%%%%%%%%%%%%%%%%%%%%%%%%%%
%%%%%%%%%%%%  HERE'S THE ABSTRACT  %%%%%%%%%%%%%%%%%%
%%%%%%%%%%%%%%%%%%%%%%%%%%%%%%%%%%%%%%%%%%%%%%%%%%%%%
\begin{abstract}
 Scanning diamond magnetometers based on the optically detected magnetic resonance of the nitrogen-vacancy centre offer very high sensitivity and non-invasive imaging capabilities when the stray fields emanating from ultrathin magnetic materials are sufficiently low  ($ < 10 \, \mathrm{mT}$). Beyond this low-field regime, the optical signal quenches and a quantitative measurement is challenging. While the field-dependent NV photoluminescence can still provide qualitative information on magnetic morphology, this operation regime remains unexplored particularly for surface magnetisation larger than $\sim 3 \, \mathrm{mA}$. Here, we introduce a multi-angle reconstruction technique (MARe) that captures the full nanoscale domain morphology in all magnetic-field regimes leading to NV photoluminescence quench. To demonstrate this, we use [Ir/Co/Pt]$_{14}$ multilayer films with surface magnetisation an order of magnitude larger than previous reports. Our approach brings non-invasive nanoscale magnetic field imaging capability to the study of a wider pool of magnetic materials and phenomena.
\end{abstract}
%%%%%%%%%%%%%%%%%%%%%%%%%%%%%%%%%%%%%%%%%%%%%%%%%%%%%
%%%%%%%%%%%%%%%%%%%%%%%%%%%%%%%%%%%%%%%%%%%%%%%%%%%%%
%%%%%%%%%%%%%%%%%%%%%%%%%%%%%%%%%%%%%%%%%%%%%%%%%%%%%

\maketitle

\section{Introduction}
 In the last decade, the negatively-charged nitrogen-vacancy (NV) centre in diamond has attracted great interest as a versatile quantum sensor for the investigations of weak-field magnetism which demands high sensitivity, nanoscale resolution and noninvasiveness.~\cite{Balasubramanian_2008, Maletinsky2012, Tetienne2014, Gross2017, Thiel2019}. In the presence of a magnetic field, the Zeeman splitting of the NV spin can be quantified by performing optically detected magnetic resonance (ODMR) measurements using laser and microwave excitation~\cite{Rondin2014}. The single-spin nature of the NV centre also ensures limited perturbation of the measured system. Further, attaching an NV-containing diamond platform on a scanning probe~\cite{Tetienne2015,Tetienne2014,Rondin2012, Maletinsky2012,Appel2016} enables scanning NV microscopy (SNVM), which allows for nanoscale noninvasive magnetic imaging. This technique features a large operating temperature range (cryogenic to room temperature) and stability in vacuum to ambient conditions~\cite{Rondin2014,Balasubramanian_2008,Degen_2008}. However, the ODMR measurements are restricted to magnetic fields below \numwunit{10}{mT} due to the field-induced quenching of the ODMR contrast, thus preventing the optical readout of the spin splitting~\cite{Dovzhenko2018,Tetienne2012,Rondin2012}. As a consequence, quantitative ODMR-based SNVM has been demonstrated mainly on magnetic textures in thin films with close to zero surface magnetisation, such as antiferromagnetic or single layer ferromagnetic materials~\cite{Rondin2012, Tetienne2014-1,Tetienne2015, kosub2017,Gross2017, Dovzhenko2018,Wornle2019,Jenkins2019,Appel2019, Sun2020,Hedrich2020,Wornle2020}.
 \begin{figure}
    \centering
    \includegraphics[width = 1 \columnwidth]{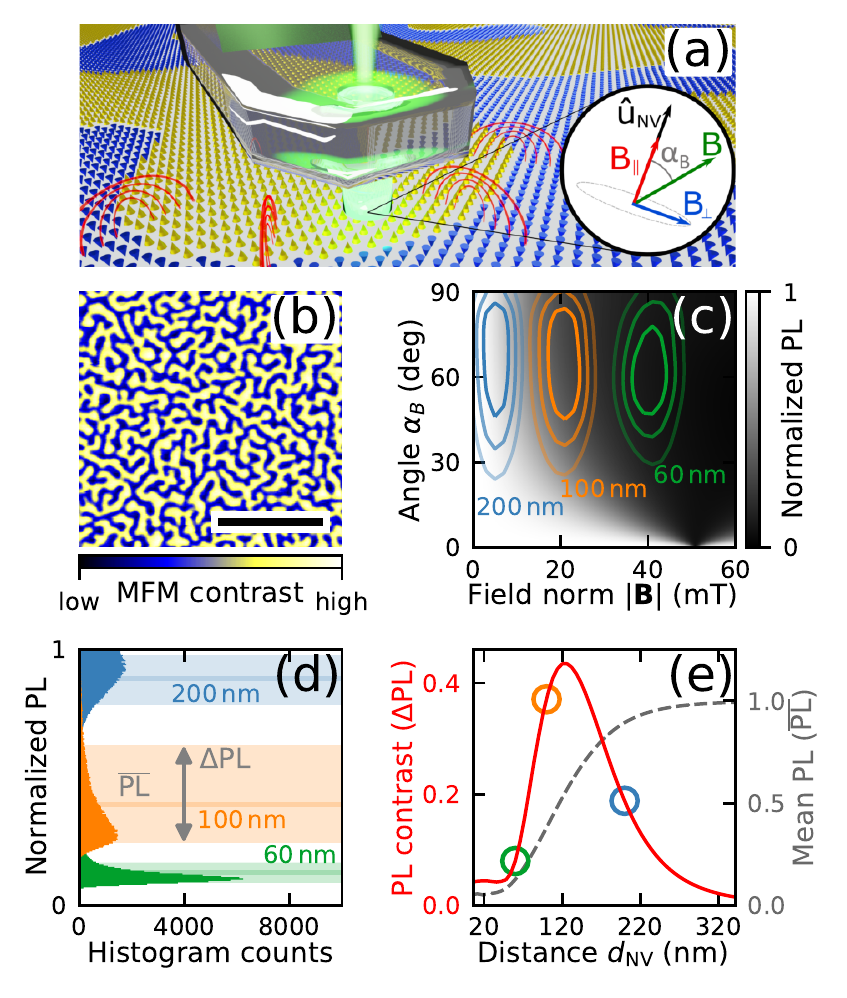}
    \caption{ \textbf{Effect of Magnetic Field Amplitude and Orientation on NV Luminescence.} 
    (a) Illustration of a diamond probe scanning over a spin texture (colored cones) with magnetic field lines across the domain boundaries (red lines). The inset is a schematic of the local magnetic field vector $\textbf{B}$ with reference to the NV quantisation axis \unv{} at the tip of the diamond probe. $\alpha_B$ indicates the angle between $\textbf{B}$ and \unv{}. (b) Labyrinth domain morphology in Ir/Co/Pt multilayer observed by MFM, exhibiting a zero-field period of \numwunit{407}{nm} (scale bar: \numwunit{3}{\mu m}).
    (c) Normalized NV luminescence defined as $(\mathrm{PL} - \mathrm{PL}_{\rm \min}) / (\mathrm{PL}_{\rm \max} - \mathrm{PL}_{\rm \min})$ as a function of $\left| \mathbf{B} \right|$  and $\alpha_B$. The corresponding distributions at various \dnv{}, obtained from simulated magnetic fields across (b), are overlaid on (c) (Contour lines). The 80th, 60th, and 40th percentiles are indicated with increasingly lighter contour lines. (d) The histograms of the simulated PL response at the three \dnv{} values, $\numwunit{60}{nm}$, $\numwunit{100}{nm}$ and $\numwunit{200}{nm}$. $\overline{\mathrm{PL}}$ indicates the mean PL, $\Delta \mathrm{PL}$ marks the difference between the 90th and 10th percentile of the PL distribution. (e) Dependence of $\Delta \mathrm{PL}$ and $\overline{\mathrm{PL}}$ on \dnv{}. The peak of $\Delta \mathrm{PL}$ marks the optimal distance for quench-based imaging for the multilayer film of (b). The colored circles correspond to the three \dnv{} values considered in panels (c) and (d).
 }
    \label{fig:figure_stray_quench}
\end{figure}

To extend the operational range beyond \numwunit{10}{mT}, the NV centre can harness the field-dependent quench of the NV photoluminescence (PL) for magnetic imaging as demonstrated recently~\cite{Gross2018,Dovzhenko2018,Akhtar2019,Rana2020}. Quench-based SNVM monitors the changes in NV PL due to the local magnetic field variation across a spin texture with the respect to the NV quantization axis. This modality also offers reduced acquisition time and enables microwave-free non-perturbative operation~\cite{wickenbrock2016, Thiel2019,Zheng2020}. The interpretation of quench-based SNVM maps can be ambiguous, because of the multiple parameters that influence PL quenching, such as NV-sample distance, NV axis orientation, sample magnetization, magnetic domain size or magnetic field noise~\cite{finco2020}. Therefore, this imaging mode has been limited to the mapping of magnetic domain morphology with surface magnetisation $I_{S} \lesssim \numwunit{3}{mA}$~\cite{Gross2018,Akhtar2019,Rana2020} (equivalent to $~\numwunit{2}{nm}$ of $\mathrm{Co}$). In this report, we reveal distinct quench-based imaging regimes, dependent on the material parameters, and introduce the Multi-Angle Reconstruction (MARe) protocol to interpret the domain morphology from quenched SNVM maps. We demonstrate MARe on [Ir/Co/Pt]$_{14}$ multilayer film with \numwunit{12}{mA} out-of-plane surface magnetisation, an order of magnitude larger than the operational limit of ODMR-based SNVM. Utilising MARe can extend the applicability of SNVM to a wider range of materials and magnetic regimes.

\section{Quench-Based Imaging in Different Regimes}
Figure~\ref{fig:figure_stray_quench}(a) illustrates our experimental setup consisting of a diamond scanning probe with an NV centre implanted close to the diamond surface at an NV-sample distance \dnv{} smaller than $\numwunit{100}{nm}$~\cite{Maletinsky2012, VanDerSar2015, Appel2016, Zhou2017}. The optical ground state of the NV centre is a spin triplet, with a quantisation axis \unv{} along one of the four crystallographic axes of the diamond lattice~\cite{Doherty2012,Doherty2013} and the lowest-energy state $\ket{m_s=0}$ is split from the $\ket{m_s = \pm 1}$ states by \numwunit{2.87}{GHz}~\cite{Doherty2011}. The local magnetic field can be decomposed into parallel ($\textbf{B}_{\parallel}$) and orthogonal ($\textbf{B}_{\perp}$) components with respect to \unv{} (Fig.~\ref{fig:figure_stray_quench}(a) insert). The $\textbf{B}_{\parallel}$ splits the $\ket{m_s = \pm 1}$ states which is measured by monitoring the ODMR~\cite{Gruber1997}. However, $\textbf{B}_{\perp}$ mixes these spin states and modifies the branching ratio of the optical transitions~\cite{Tetienne2012}. This results in the quenching of the NV PL and the suppression of the ODMR contrast (Supplemental Material~\ref{sec:photodynamics}), restricting quantitative ODMR-based imaging to below $\sim\numwunit{10}{mT}$~\cite{Tetienne2012,Dovzhenko2018}.
\begin{figure*}
    \centering
    \includegraphics[width = 0.7\textwidth]{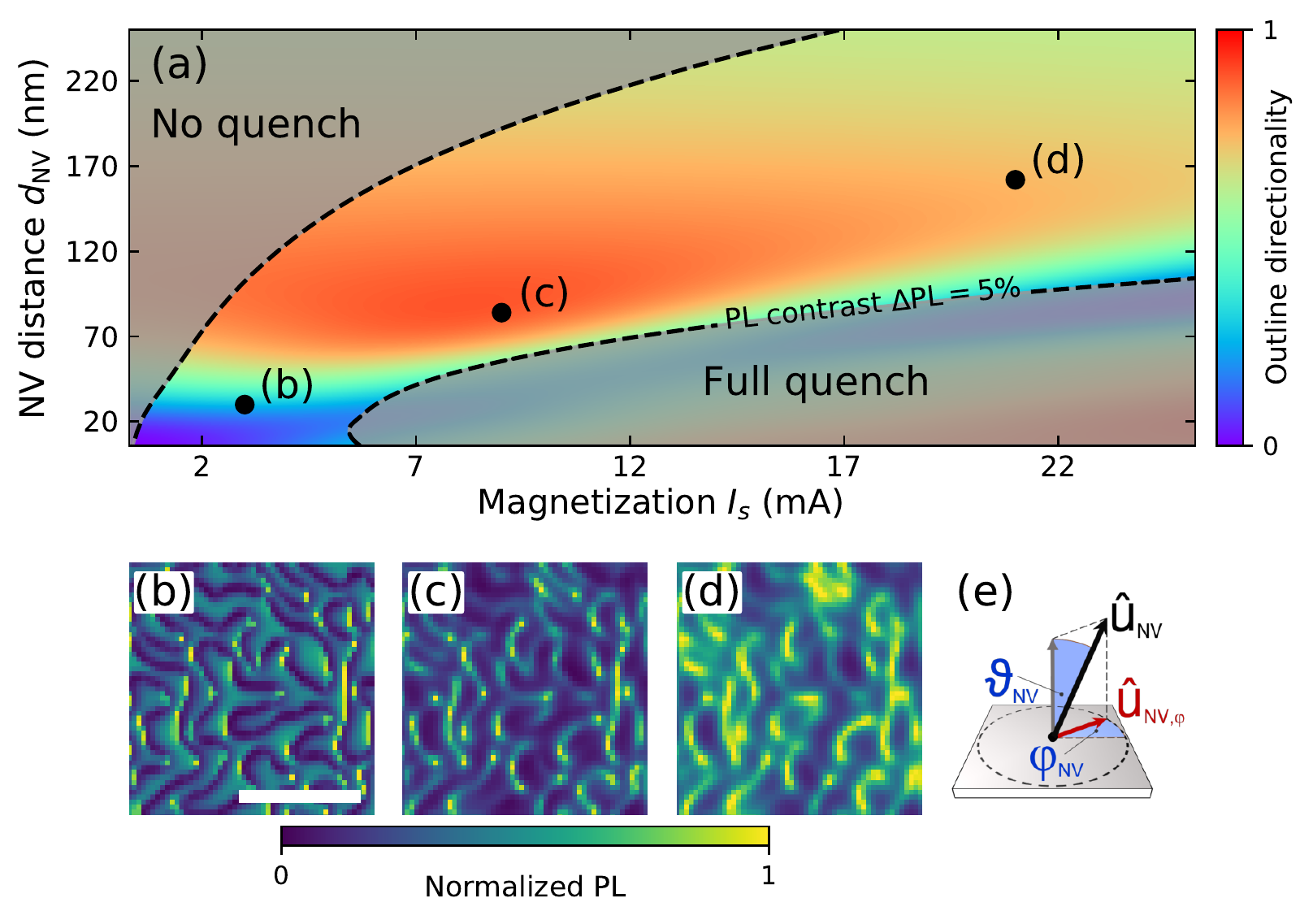}
    \caption{ \textbf{Quench-based SNVM Imaging Regimes.} (a) Different regimes of quench-based imaging as a function of NV-sample distance (\dnv{}) and surface magnetisation ($I_s$), based on simulated quench images. Little to no domain morphological information is captured in the \textit{No Quench} (left greyed area) and \textit{Full Quench} (right greyed area) regimes where the PL map is predominantly bright or dark, respectively. In the \textit{Partial Quench} regime (area bounded by dashed lines), field variations are mapped to PL changes resulting in (b-d) quench images with features indicative of domain boundaries (scale bar: \numwunit{1}{\mu m}). Domain boundaries appear as dark isotropic PL features (low directionality) for smaller \dnv{} and $I_s$ (b), and as directional bright features (high directionality) at larger \dnv{} and $I_s$ (c-d). The orientation of the directionality depends on \unvphi{} which is the NV axis \unv{}, projected on the sample surface. The dashed lines indicate the contour lines for 5\% map contrast. (e) Illustration depicting the NV axis \unv{}, the tilt angle \thetanv{} from the normal to the sample surface, and the projection of \unv{} in the sample plane, \unvphi{}. The angle \phinv{} is the angle between \unvphi{} and the reference axis within the sample plane. For panels (a-d), $\thetanv{} = 54.7^\circ$ and $\phinv{} = 0^\circ$.
    }
    \label{fig:figure_im_regimes}
\end{figure*}

Quench-based SNVM generates a PL intensity map, where regions with strong $\textbf{B}_{\perp}$ component appear darker. In the limit of modest surface magnetisation and small NV-sample distance \dnv{}, the domain boundaries appear dark producing faithful magnetic domain morphology maps. Therefore, demonstrations are limited to single or bilayer thin film systems with surface magnetisation $I_s \lesssim \numwunit{3}{mA}$ ~\cite{Gross2018,Akhtar2019,Rana2020,Dovzhenko2018,Rondin2012}. Outside this regime, the complex interplay between \dnv{} and $I_s$, as well as the morphology lenghtscale, on the NV PL obfuscates the straightforward correspondence of dark regions to domain boundaries. Therefore, a systematic understanding of quench-based SNVM response is necessary to retrieve the domain morphology of a magnetic material. To do this, we first simulate the \dnv{} dependence of quench-based SNVM for a known magnetic structure.

Our study involves [Ir(\numwunit{1}{nm})/Co(1)/Pt(1)]$_{14}$  magnetic multilayer, a room-temperature skyrmion platform with an out-of-plane anisotropy and $I_s= \numwunit{12}{mA}$  (Supplemental Material~\ref{sec:mat_prop_and_prep}) -- an order of magnitude larger than systems studied previously with SNVM. Further, the ambient stability of the nanoscale spin textures~\cite{Moreau-Luchaire2016, soumyanarayanan2017} allows us to correlate the quench-based SNVM images with MFM measurements~\cite{kazakova2019}. Figure~\ref{fig:figure_stray_quench}(b) presents an MFM image of this film, exhibiting a labyrinth domain morphology with a zero-field period of \numwunit{407}{nm}. Figure~\ref{fig:figure_stray_quench}(c) presents a grey-scale map of normalised PL intensity simulated as a function of field amplitude $\left|\textbf{B}\right|$ and field angle $\alpha_B$ with respect to the NV axis \unv{}. To understand how the stray field distribution of the domain morphology affects the NV PL at various \dnv{}, we simulate the volumetric field distribution from the MFM map in Figure~\ref{fig:figure_stray_quench}(b) using the micromagnetics package mumax$^3$~\cite{mumax} (Supplemental Material~\ref{sec:micromagnetics}). On Figure~\ref{fig:figure_stray_quench}(c), we overlay the corresponding $\left|\textbf{B}\right|$~-~$\alpha_B$ distributions of the magnetic field at three different \dnv{}, at $\numwunit{60}{nm}$ (green contours), $\numwunit{100}{nm}$ (orange) and $\numwunit{200}{nm}$ (blue). (Supplemental Material~\ref{sec:quenching_regimes}). At $\dnv{}=\numwunit{60}{nm}$ ($\numwunit{200}{nm}$), the NV PL remains uniformly quenched (unaffected) for the majority of the field distribution, while $\numwunit{100}{nm}$ \dnv{} results in strong PL variation. Figure~\ref{fig:figure_stray_quench}(d) clearly highlights these NV PL variations $\Delta \mathrm{PL}$ via the corresponding histograms at $\dnv{}=60,\, 100$ and \numwunit{200}{nm}. Figure~\ref{fig:figure_stray_quench}(e) presents the $\Delta \mathrm{PL}$ -- calculated as the difference between the 90th and 10th percentile of the NV PL distribution -- as a function of \dnv{} (solid red curve) alongside the mean PL (dashed grey curve).
\begin{figure}
    \centering
    \includegraphics[width = \columnwidth]{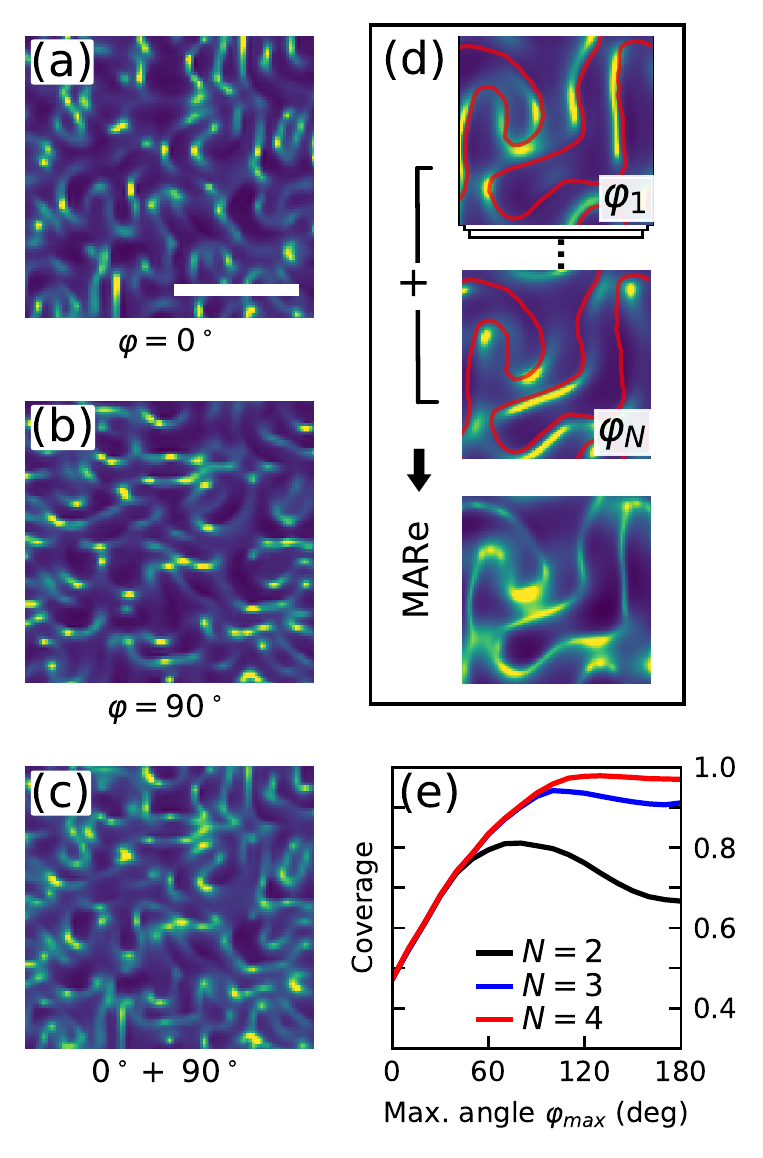}
    \caption{ \textbf{Directional Quench Imaging and morphology reconstruction.} (a) Simulated quenched PL map based on spin texture in Figure~\ref{fig:figure_stray_quench}b, with $\thetanv{} = 54.7^\circ$, $\phinv{} = 0^\circ$ and (b) simulated quenched PL map in the same area but with the NV rotated $90^\circ$ in the sample plane ($\phinv{} = 90^\circ$). Both maps are simulated at a NV-sample distance $\dnv{} = \numwunit{77}{nm}$ and surface magnetisation $I_{s} = \numwunit{12}{mA}$ (scale bar: \numwunit{1}{\mu m}). (c) Reconstructed image obtained by summing (a) and (b). (d) \textbf{M}ulti-\textbf{A}ngle \textbf{Re}construction (MARe) illustrating the domain morphology acquisition based on multiple \textit{N} images at various \phinv{}. (e) Coverage of domain boundaries given as function of $N$ and $\varphi_{max}$. $N$ is the number of quench images involved in the reconstruction, and are obtained over a range of $\phinv{}$ ($0^\circ$ to $\varphi_{max}$) spaced by $\Delta\phinv{} = \varphi_{max}/(N-1)$. The reconstruction with $N=4$ images yields the largest coverage, which saturates above $\varphi_{max} \simeq 120^\circ$.}
    \label{fig:mfm_dirquench}
\end{figure}

To assess the operational regime of quench-based SNVM, we need to consider further the interplay between $I_s$ and \dnv{}. As shown in Figure~\ref{fig:figure_im_regimes}(a), quench-based SNVM can be categorised into different regimes. The combination of large \dnv{} and small $I_s$ (small \dnv{} and large $I_s$) results in predominantly bright (quenched) PL maps. In both the \textit{No Quench}  and the \textit{Full Quench} regimes, the lack of PL variation $\Delta \mathrm{PL}$ implies that little to no morphological information of the underlying spin textures is captured. In contrast, quench-based SNVM is feasible in the \textit{Partial Quench} regime (area bounded by dotted lines in Figure~\ref{fig:figure_im_regimes}(a)) for a limited range of $I_s$ and \dnv{} combinations. While the \textit{Partial Quench} regime gives a large  $\Delta \mathrm{PL}$, which is desirable for quench-based SNVM, the resultant PL maps over an identical spin texture can vary dramatically across this regime. To highlight this, we simulated quench-based SNVM maps of the same area in the multilayer film using three different combinations of $I_s$ and \dnv{} (Fig.~\ref{fig:figure_im_regimes}(b-d)). In general, we observe an evolution from dark, isotropic features at lower $I_s$ and \dnv{} to bright, directional features at higher $I_s$ and \dnv{} due to competing magnetic field contributions above domains and domain boundaries. At lower $I_s$ and \dnv{} (blue region in Figure~\ref{fig:figure_im_regimes}(a)), the quench image appears as a uniform bright background with isotropic dark outlines (Fig.~\ref{fig:figure_im_regimes}(b)). This is a result of the strong magnetic field localised at the domain boundaries which quenches the NV. The NV quench images reported to-date lie in this region of the parameter space~\cite{Gross2018, Rondin2012, Akhtar2019, Rana2020} (Supplemental Material~\ref{sec:quenching_regimes}).

\begin{figure*}
    \centering
    \includegraphics[width = 0.7\textwidth]{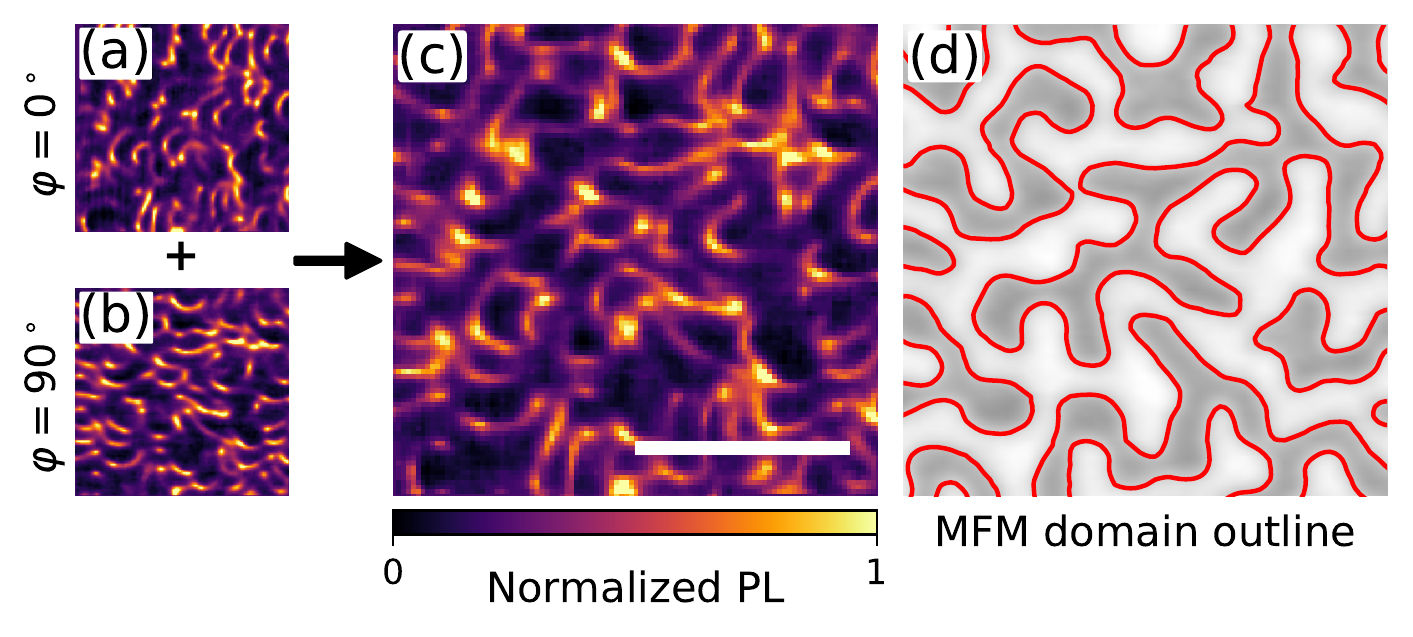}
    \caption{\textbf{Experimental verification of Multi-angle Reconstruction of Domain Morphology.}
    Experimental quenching map of the same area of Figure~\ref{fig:mfm_dirquench}(a, b) with $\thetanv{}=60\pm2^\circ$ and (a) with $\varphi = 0^\circ$ and (b) with $\varphi  = 90^\circ$. The two images are combined to give
    (c) the reconstructed domain morphological map with $N=2$. (scale bar: \numwunit{1}{\mu m}). (d) Binarized and magnified image of Figure~\ref{fig:figure_stray_quench}(b) covering the same area in (a, b and c), with domain boundaries marked in red.} 
    \label{fig:reconstruct}
\end{figure*}

For combinations of larger $I_s$ and \dnv{} values (orange region in Figure~\ref{fig:figure_im_regimes}(a)), the quench maps generate strikingly different images: panels (c) and (d) capture highly directional bright and segmented features along the domain boundaries. In this case, strong off-axis magnetic field above domains and domain boundaries results in a predominantly dark PL map. However due to large gradients localised at domain boundaries, there are instances where the field is aligned closer to \unv{}. This occurs across portions of domain boundaries orthogonal to the projection of \unv{} in the sample plane (\unvphi{}), resulting in directional bright features for panels (c) and (d), highly dependent on the NV equatorial angle \phinv{} (Fig.~\ref{fig:figure_im_regimes}(e)). Notably, this directional behaviour occurs over a significantly larger parameter space of the \textit{Partial Quench} regime, well beyond that of panel (a), and the trend remains valid for different domain periodicity (Supplemental Material~\ref{sec:quenching_regimes}). It is worth emphasising here that magnetic materials with $I_s$ larger than $\sim\numwunit{3}{mA}$ would inevitably constrain quench-based SNVM to the directional region of Figure~\ref{fig:figure_im_regimes}(a). Therefore, a protocol that relates these images with the actual magnetic domain morphology is necessary in order to extend the operation regime of quench-based SNVM for non-perturbative investigations of such materials. 

\section{Reconstruction of domain morphology - MAR\MakeLowercase{e}}
To reflect the role of \phinv{} in quench-based SNVM, we simulate two quench images for $\phinv{} = 0^\circ$ and $\phinv{} = 90^\circ$, displayed in Figure~\ref{fig:mfm_dirquench}(a) and~\ref{fig:mfm_dirquench}(b), respectively. We set \dnv{} = \numwunit{77}{nm}, $\thetanv{} = 54.7^\circ$, and $\sim\numwunit{12}{mA}$ surface magnetisation (Supplementary Material~\ref{sec:dnv_estimate}) to reflect our experimental measurements. The images for both \phinv{} orientations show directional segments revealing some features of the domain morphology, but more importantly these segments are complementary. Therefore, while an image at a given \phinv{} remains incomplete, images obtained at multiple \phinv{} values can collectively give a significantly better coverage of the underlying domain morphology, which is the essence of the proposed imaging protocol. Multi-Angle Reconstruction protocol (MARe) harnesses the \phinv{} dependence of PL features to build a composite map enabling morphological imaging further into the \textit{Partial Quench} regime, i.e. in strong-field conditions.

The overlapping features in the PL maps obtained at different \phinv{}, e.g. $0^\circ$ and $90^\circ$ as in panels (a) and (b) of Figure~\ref{fig:mfm_dirquench} allow us to perform an initial image registration to compensate for the domain outline shift caused by $\thetanv{} \neq 0^\circ$ (Supplemental Material~\ref{sec:111_nv}). Subsequently, the maps are normalised and summed to yield a MARe image, as displayed in Figure~\ref{fig:mfm_dirquench}(c), revealing a larger fraction of the domain boundaries with just two values of \phinv. To quantify the domain boundary coverage, we integrate the product of the domain outlines from the MFM image (Fig.~\ref{fig:figure_stray_quench}(b)) with the binarized MARe image. In order to maximise the fraction of domain boundaries covered by the protocol, we consider $N \ge 2$ images taken at different \phinv{} values ranging from $0^\circ$ to $\phinv^{\rm max}$ spaced equally by $\Delta\phinv{}=\phinv^{\rm max}/(N-1)$. Figure~\ref{fig:mfm_dirquench}(d) illustrates the MARe scheme for $N=4$ and $\varphi_{\rm max} = 120^\circ$, which corresponds to 4 quench-based SNVM images with each obtained at $40^\circ$ relative angle. The corresponding MARe image  clearly captures an increased fraction of the domain morphology. 

Figure~\ref{fig:mfm_dirquench}(e) presents the calculated fraction of domain boundary coverage for MARe with $N=2,3$ and $4$ (black, blue and red curve). For $N=2$ ($3$) the maximum coverage reaches $81\%$ ($94\%$) at $\varphi_{\rm max} = 80^\circ$ ($100^\circ$). Extending MARe to $N=4$ further improves the coverage reaching a maximum of $\sim 98\%$. This shows that even for $N\le 4$, the MARe protocol is capable of recovering the domain morphology with near-unity coverage.

Figure~\ref{fig:reconstruct} presents our experimental demonstration of domain morphology mapping using MARe on the [Ir/Co/Pt]$_{14}$ multilayer. To obtain quench-based SNVM images we use a (100) diamond probe containing a single NV centre with $\thetanv{} = \numwunit{60 \pm 2}{^\circ}$ and $\dnv{} = \numwunit{77 \pm 3}{nm}$ (Supplementary Material~\ref{sec:dnv_estimate}). The combination of the \dnv{} ($\numwunit{77}{nm}$) value and  $I_s~(\numwunit{12}{mA})$ of the [Ir/Co/Pt]$_{14}$ multilayer yields  directional quench images according to Figure~\ref{fig:figure_im_regimes}(a). Figures~\ref{fig:reconstruct}(a) and \ref{fig:reconstruct}(b) show experimental quench images acquired at $\phinv{} = 0^\circ$ and $\phinv{} = 90^\circ$, respectively, on the same area used for simulating Figure~\ref{fig:mfm_dirquench}(a) and ~\ref{fig:mfm_dirquench}(b) (Supplemental Material~\ref{sec:mat_prop_and_prep}). The domain boundary coverage of each of these images is $60\substack{ +13 \\ -11}\%$, in line with the simulations and there is good agreement between the simulated and the measured images for both orientations (Supplemental Material~\ref{sec:domain_coverage}). Figure~\ref{fig:reconstruct}(c) is the corresponding $N=2$ MARe image showing matching bright features with the highlighted domain boundaries of the binarised MFM image displayed in Figure~\ref{fig:reconstruct}(d). The experimentally achieved domain boundary coverage is improved to $71\substack{ +12 \\ -15}\%$ -- an enhancement beyond the single frame coverage of $\sim 60\%$. The deviation from the simulated coverage is due to the nonlinearity of the experimental map, as well as image thresholding and registration operations (see Supplementary Material~\ref{sec:im_reconstruct}).  Another reason for this deviation might be due to perturbations of the domain morphology induced by MFM scanning. As the experimental protocol includes MFM scans performed before and after each quench-based SNVM map, we do observe local perturbations due to MFM that could potentially lead to deviations from the unperturbed images captured by quench-based SNVM (see Supplemental Material~\ref{sec:sample_perturbation}). Nonetheless, the experimental demonstration of MARe extends the operational range of non-invasive quench-based SNVM into the \textit{Partial Quench} regime. 

\section{Outlook}
%\subsection{Summary} 
Our work methodically evaluates quench-based SNVM in terms of characteristic NV and magnetic material properties. We establish a predictive scheme involving MFM, micromagnetics and NV photodynamics simulations, which yields images in excellent agreement with experimentally acquired data. We find two regimes of quench imaging where morphological information is captured. The first regime corresponds to mostly bright PL maps with dark outlines tracing the domain boundaries, which corresponds to materials of low magnetisation ($I_s < \numwunit{3}{mA}$). The second regime, which has not been reported to-date, results in PL maps with directional segmented features with strong \unvphi{} dependence. We established a multi-angle reconstruction scheme, herein named as MARe, to enable domain morphology mapping with near-unity coverage for the second regime. The experimentally validated MARe protocol extends quench-based SNVM imaging of out-of-plane spin textures to magnetic systems with $I_s > \numwunit{3}{mA}$. Furthermore, the scheme to identify the imaging regimes can be generalized to complex magnetic textures, thus enabling the forecast of the attainable SNVM modes. We anticipate that these insights, alongside tools developed for prediction, interpretation and reconstruction, will stimulate the adoption of quench-based SNVM as a non-perturbative nanoscale magnetometry to a wider pool of materials, thereby furthering the development of quantitative quench-based SNVM imaging.

\section{Acknowledgements}
This work was performed at the Cambridge Nanoscale Sensing and Imaging Suite (CANSIS), part of the Cambridge Henry Royce Institute, EPSRC grant EP/P024947/1. We further acknowledge funding from EPSRC QUES2T (EP/N015118/1) and from the Betty and Gordon Moore Foundation.
This work was also supported by the Faraday Institution (FIRG01) and by the SpOT-LITE programme (Grant Nos. A1818g0042, A18A6b0057), funded by Singapore’s RIE2020 Initiatives. A. K. C. Tan acknowledges funding from A*STAR, Singapore. B. Vindolet acknowledges support by a PhD research Grant of D\'el\'egation G\'en\'erale de l’Armement.
J.-F. Roch thanks Churchill College and the French Embassy in the UK for supporting his stay at the Cavendish Laboratory.

%%%%%%%%%%%%%%%%%%%%%%%%%%%%%%%%%%%%%%%%%%%%%%%%%%%%%%%%%%%%%%%%%%%%%%%%%%%%%
%%%%%%%%%%%%%%%%%%%%%%%%%%%%%%%%%%%%%%%%%%%%%%%%%%%%%%%%%%%%%%%%%%%%%%%%%%%%%
%%%%%%%%%%%%%%%%%%%%%%%%%%%%%%%%%%%%%%%%%%%%%%%%%%%%%%%%%%%%%%%%%%%%%%%%%%%%%
\newpage\hbox{}\thispagestyle{empty}\newpage
\onecolumngrid
\appendix

\makeatletter 
\renewcommand{\thefigure}{S\@arabic\c@figure}
\makeatother

\makeatletter 
\renewcommand{\thetable}{S\@arabic\c@table}
\makeatother

\makeatletter 
\renewcommand{\thesection}{\Alph{section}}
\makeatother

\section{Simulation of the NV photodynamics} \label{sec:photodynamics}
To capture the photodynamics of the NV centre, we use of a seven-state model which includes the ground-state and excited-state fine structure of the NV centre (Fig.~\ref{fig:Sfig_photodyn_levels}). The strain splitting is $E_{\rm gs} = E_{\rm es} \approx 0$, where the subscripts gs and es indicate the optical ground state and the optical excited state, respectively. At zero-field, the levels $\ket{i}$ with $i = 0, 1, 2$ are split by $D_{\rm gs}=\numwunit{2.87}{GHz}$ in the optical ground state while the levels of the excited state $\ket{i}$, $i = 3, 4, 5$, are split by the excited state zero-field splitting $D_{\rm es} = \numwunit{1.42}{GHz}$. The transition rates from level $\ket{i}$ to the level $\ket{j}$ are denoted as $\gamma_{ij}$. The decay rates are defined as in the work by Tetienne et al.~\cite{Tetienne2012}: we assume $\gamma_{30} = \gamma_{41} = \gamma_{52}= \gamma_r$, $\gamma_{46} =\gamma_{56}$, and $\gamma_{61} = \gamma_{62}$. The spin non-conserving transitions from the excited state are assumed to be forbidden. Optical excitation pumps the ground state populations to the excited state but stimulated emission is neglected, the laser being off-resonant and the vibrational relaxation decay time being short. The values used for the numerical simulations are taken from the works by Robledo et al. and Tetienne et al.~\cite{Robledo2011a,Tetienne2012} (Tab.~\ref{tab:decayrates}).
Within the assumption of Markovian noise:
\begin{equation}
	\timeder{\rho(t)} = -\frac{i}{\hbar} \left[ \mathscr{H}, \rho \right] - \frac{1}{2} \sum_{k=0}^{m} \left( L_k^\dagger L_k \rho + \rho L_k^\dagger L_k \right) + \sum_{k=0}^{m} L_k \rho L_k^\dagger \; \,
	\label{eq:lindbladeq}
\end{equation}
where $\mathscr{H}$ is the magnetic-field dependent Hamiltonian describing the seven-state system, $\rho$ is the density operator and $L_k$ are the Kraus operators which describe the $m$ photon emission or absorption processes. We work in the approximation of microwave excitation rate weaker than the laser pumping, hence $T_2^*$ dephasing is neglected. The laser pump is described as an incoherent absorption process. The Kraus operators then can either take the form:
\begin{equation}
L_k^{abs} = \sqrt{\gamma_{ji}} \, \ketbra{i}{j}, \; i=\left(3,4,5\right), j = \left(0,1,2 \right) \; \,
\end{equation}
or
\begin{equation}
L_k^{em} = \sqrt{\gamma_{ij}} \, \ketbra{j}{i}, \; i=\left(3,4,5,6\right), \, j = \left(0,1,2 \right) \; .
\end{equation}
Extra Kraus operators can be added if incoherent microwave driving is included in the model:
\begin{equation}
L_k^{mw} = \sqrt{\gamma_{ij}^{mw}} \ketbra{i}{j}, \; i,j=\left(0,1,2\right), i \neq j \; .
\end{equation}
The steady-state PL rate is proportional to the sum of the steady-state populations in the excited state:
\begin{equation}
\Gamma_{\mathrm{PL}} \propto \sum_{i=3}^{5} \rho_{ii} \, .
\end{equation}

\begin{minipage}{0.43\textwidth}
    \centering
    \begin{tabular}{||c|c||}
    		\Xhline{3\arrayrulewidth}
    		& Decay rate (MHz) \\
    		\Xhline{3\arrayrulewidth}
    		$\gamma_r$ &  65 \\ 
    		\hline 
    		$\gamma_{36}$ &  11 \\ 
    		\hline 
    		$\gamma_{46} = \gamma_{56}$ & 80  \\ 
    		\hline 
    		$\gamma_{60}$ & 3 \\ 
    		\hline 
    		$\gamma_{61} = \gamma_{62}$ & 3 \\ 
    		\hline 
    	\end{tabular}
    	\captionof{table}{\textbf{Photodynamics parameters.} The previously reported~\cite{Robledo2011a,Tetienne2012} decay rates used in the 7-state model for the NV magnetic field-dependent photodynamics.}
    	\label{tab:decayrates}
\end{minipage}
\hfill
\begin{minipage}{0.49\textwidth}
    \centering
    \includegraphics[width = \textwidth]{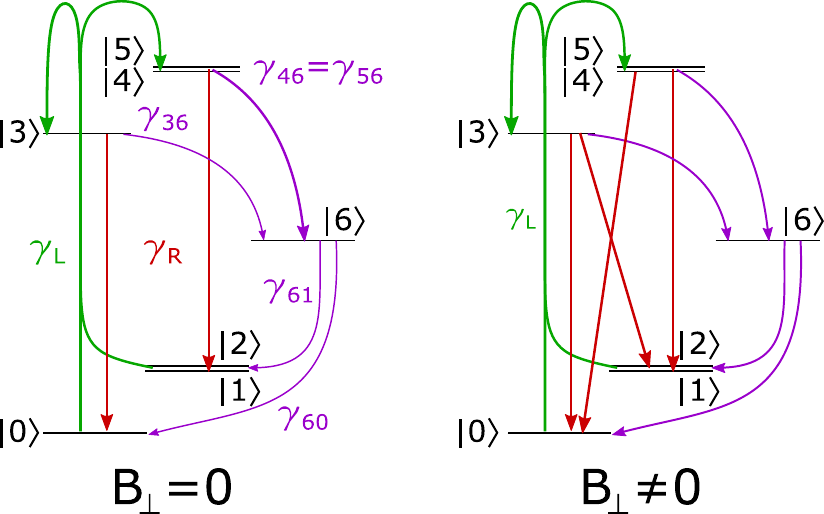}
	\captionof{figure}{\textbf{Schematics of the NV seven-level system.} Seven-level system used to capture the NV photodynamics, for an arbitrary magnetic field. In general, off-axis magnetic fields couple the zero-field eigenstates and allow for spin-flip transitions which modify the zero-field photodynamics. Green lines represent laser excitation, red lines optical decay and purple lines non-radiative decay.}
	\label{fig:Sfig_photodyn_levels}
\end{minipage}
\vspace{12pt}
\begin{figure}
    \centering
    \includegraphics{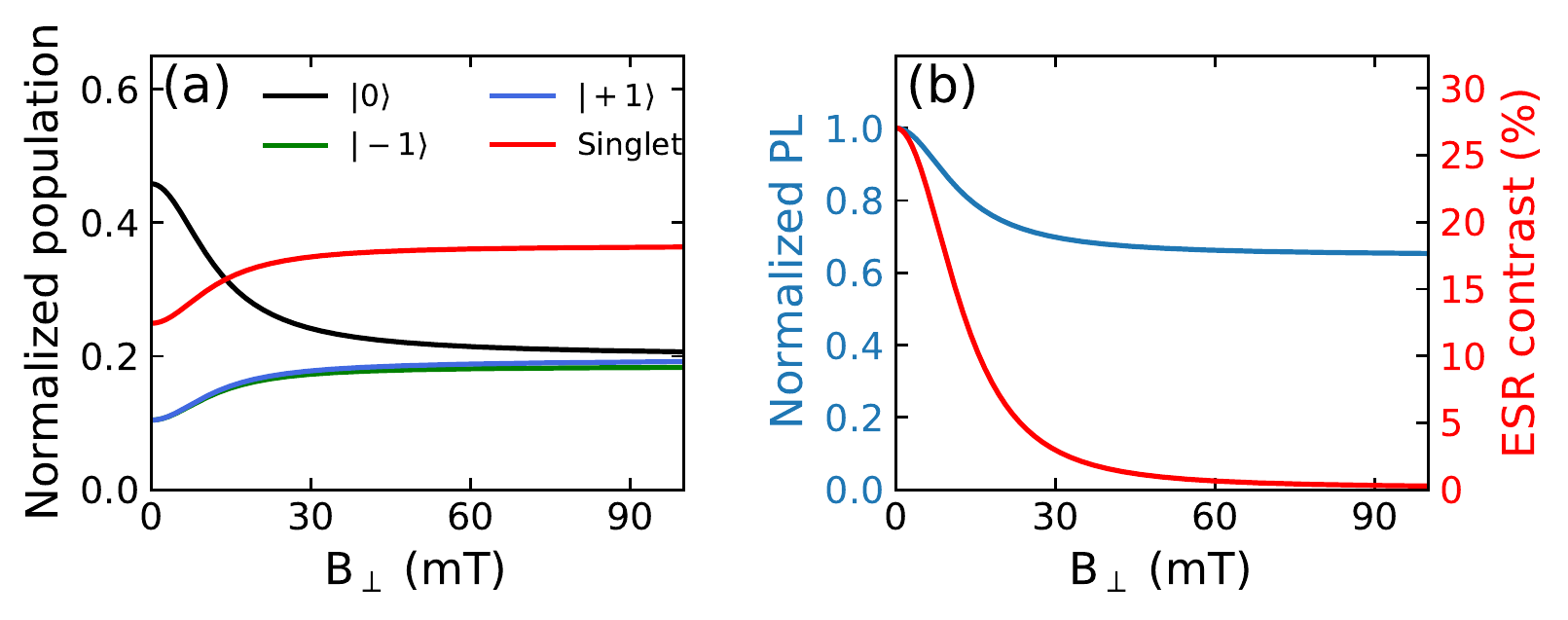}
    \caption{\textbf{Magnetic field-dependent NV photodynamics.} (a) Changes in steady state population under continuous green excitation of the triplet ground state and singlet state of an NV as a function of a magnetic field, $B_\perp$, orthogonal to the NV axis \unv{}. (b) The corresponding quench response of the NV PL (blue curve) with increasing $B_\perp$ due to larger shelving state population (shown in (a)). ODMR contrast (red curve) is also reduced due to the decrease in population difference between $\ket{0}$ and $\ket{\pm 1}$ (shown in (a)).}
    \label{fig:Sfig_thepops_contrasts}
\end{figure}

Magnetic field components orthogonal to \unv{} (off-axis) couple the different spin states, modifying the branching ratio of the transitions~\cite{Tetienne2012} and altering the steady-state populations of the levels (Fig.~\ref{fig:Sfig_thepops_contrasts}(a)). On one hand, this leads to a reduction of the ESR contrast (Fig.~\ref{fig:Sfig_thepops_contrasts}c), because of the reduced population difference between the $\ket{0}$ and $\ket{\pm 1}$ levels (Fig.~\ref{fig:Sfig_thepops_contrasts}(b)). On the other hand, this leads to the quenching of the PL~\cite{Rondin2012,Tetienne2012,Stefan2020}, due to a larger population getting trapped in the singlet state (Fig.~\ref{fig:Sfig_thepops_contrasts}(b)). This effect leads to a trade-off between magnetic field amplitude ($\propto 1/ \dnv{}$) and spatial resolution ($\propto \dnv{}$) when imaging small spin textures.

\section{Material Properties and Preparations} \label{sec:mat_prop_and_prep}
The multilayer stack of [Ir($\numwunit{1}{nm}$)/ Co(1)/ Pt(1)]$_{14}$ were deposited on thermally oxidised Si wafers by DC magnetron sputtering. Additional fabrication information is found in previous studies~\cite{soumyanarayanan2017}. Relevant properties of the Ir/Co/Pt stack are shown in Table~\ref{table:material_properties}.
The surface magnetisation $I_s$ is given by $M_s\cdot t_{\rm eff}$, where $t_{\rm eff}$ is the  effective magnetic thickness which is the number of repetition multiplied by the thickness of the magnetic layer. In this case, $t_{\rm eff}$=14\,nm, and hence $I_{s}$=12.3\,mA (Table.~\ref{table:material_systems}). The $I_s$ of various systems studied with quenched SNVM is given in Table~\ref{table:material_systems} for comparison. The zero-field magnetic domains are stabilised by demagnetising the sample. This results in labyrinth morphology with a period, $P$=407\,nm (Fig.~\ref{fig:Sfig_MFM_threshold}(a)). 

\begin{figure}
    \centering
    \includegraphics[width = 0.6\textwidth]{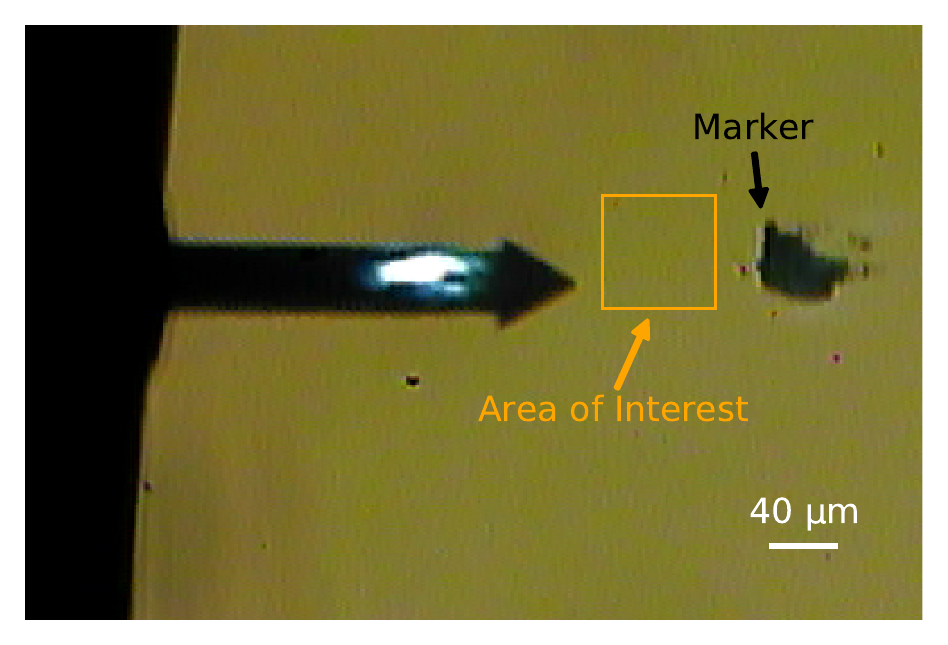}
    \caption{\textbf{Marked Sample.} Microscopic image of a marked area of the sample surface with a MFM probe in view. The marking is achieved using a wirebonding tip, and the area of interest probed by quenched SNVM, micromagnetics and MFM in the main text is highlighted in yellow.}
    \label{fig:Sfig_marker}
\end{figure}

The sample is marked with a wirebonder (Fig.~\ref{fig:Sfig_marker}) which allows us to image the same area of interest (yellow box in Fig.~\ref{fig:Sfig_marker}) using two techniques (SNVM and MFM) on separate platforms. MFM is always carried out before and after quenched SNVM, to ensure that the morphology of the probed area remains identifiable and the features are largely unchanged.

\newcolumntype{M}[1]{>{\centering\arraybackslash}m{#1}}
\renewcommand{\arraystretch}{2}
\begin{minipage}{0.45\textwidth}
\centering
    \begin{tabular}{  ||M{2cm}||M{2cm}||M{2cm}|| } 
        \Xhline{3\arrayrulewidth}
        \makecell{$\mathbf{M_s}$ \\ $(\mathrm{MA/m})$} & \makecell{$\mathbf{K_{eff}}$\\$(\mathrm{MJ/m^3})$} & \makecell{$\mathbf{D}$\\$(\mathrm{mJ/m^2})$} \\
        \Xhline{3\arrayrulewidth}
        0.881 & 0.474 & 1.25\\
        \hline
    \end{tabular}
    \captionof{table}{\textbf{Material Properties.} The saturation magnetisation $M_s$, effective anisotropy $K_{eff}$ and DMI strength $D$ of [Ir/Co/Pt]$_{14}$ film.}
    \label{table:material_properties}
\end{minipage}
\hfill
\begin{minipage}{0.45\textwidth}
\centering
        \begin{tabular}{ ||M{4cm}|M{2cm}||} 
            \Xhline{3\arrayrulewidth}
            \textbf{Material System} & \makecell{$\mathbf{I_s}$ $(\mathrm{mA})$} \\
            \Xhline{3\arrayrulewidth}
            $14\times$ Ir/Co/Pt & $12.3$ \\
            \hline
            \makecell{Pt/CFA/MgO/Ta~\cite{Akhtar2019}\\\footnotesize{CFA: Co$_2$FeAl}} & 1.8\\
            \hline
            \makecell{Pt/FM/Au/FM/Pt~\cite{Gross2018}\\\footnotesize{FM: Ni/Co/Ni}} & 2.6\\
            \hline
            \makecell{Pt/Co/NiFe/IrMn~\cite{Rana2020}} & 1.7\\
            \hline
        \end{tabular}
         \captionof{table}{ \textbf{Material Systems.} The surface magnetisation $I_s$ of various systems studied with quenched SNVM compared to [Ir/Co/Pt]$_{14}$. }
        \label{table:material_systems}
\end{minipage}

\newpage
\section{Micromagnetic simulations} \label{sec:micromagnetics}
\begin{figure}
    \centering
    \includegraphics[width = \textwidth]{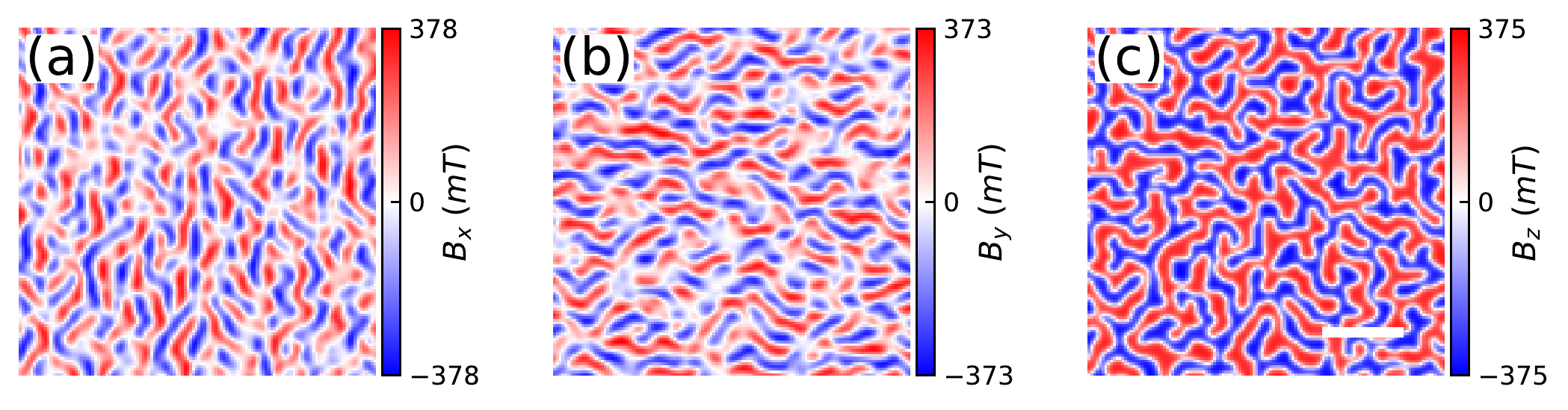}
    \caption{\textbf{Simulated Magnetic Field.} (a-c) Magnetic field components $B_x$, $B_y$, $B_z$, at $\dnv=\numwunit{77}{nm}$ above the sample surface, simulated based on the magnetisation distribution in Figure~\ref{fig:Sfig_MFM_threshold}(b). (Scale bar: \numwunit{1}{\mu m})}
    \label{fig:Sfig_Bfield}
\end{figure}

\begin{figure}
    \centering
    \includegraphics[width = \textwidth]{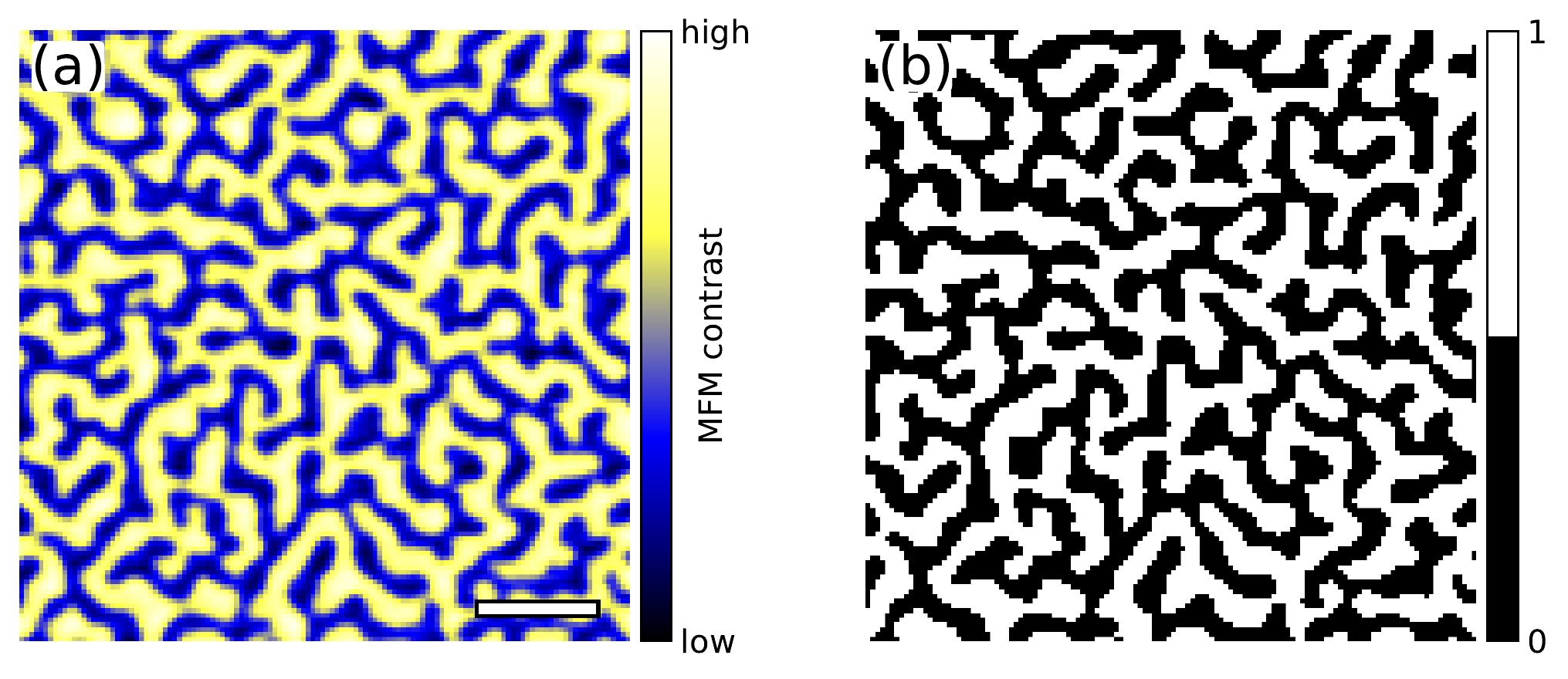}
    \caption{\textbf{Image Thresholding.} (a) MFM image of sample surface (highlighted in Figure~\ref{fig:Sfig_marker}) showing labyrinth domains at zero field. (b) Corresponding binary image after thresholding process, yielding up/down magnetisation used for simulations in Figure~\ref{fig:Sfig_Bfield}. (Scale bar: \numwunit{1}{\mu m})}
    \label{fig:Sfig_MFM_threshold}
\end{figure}

The magnetic field above the spin texture was obtained via Mumax3 simulations. For the study of quenched imaging in various regimes (Fig. 2 in main text), The multilayer film is modelled using the effective medium method~\cite{woo2016} so as to reduce computation resources. The simulation grid consists of $256 \times 256 \times 128$ cells spanning $\numwunit{10}{\mu m} \times \numwunit{10}{\mu m} \times \numwunit{384}{nm}$ (cell size: $\approx \numwunit{39}{nm} \times \numwunit{39}{nm} \times \numwunit{3}{nm}$). The first 14 layers are modelled  with an effective saturation magnetisation $M_{\rm eff} = M_s/3$ and the volume above as non-magnetic spacers.
The simulation is further refined for comparison with experiments (Fig. 3 and 4 in main text) with each cell layer corresponding to $\numwunit{1}{nm}$ of Ir, Co, or Pt. Maintaining the same grid size, this reduces the total simulated height to $\numwunit{128}{nm}$. Similarly, the Pt, Ir layer and the volume above the multilayer film are modelled as non-magnetic spacers. Differing from the effective medium model, the Co layer has the experimentally obtained magnetisation $M_s$. 
In both cases, the simulated non-magnetic volume above the multilayer film allows us to retrieve the magnetostatic field environment above the spin texture (Fig.~\ref{fig:Sfig_Bfield}) via Mumax3. The magnetisation distribution used in the simulation is based on segmenting a MFM image into up and down domains by image thresholding (Fig.~\ref{fig:Sfig_MFM_threshold}).

\newpage
\section{Analysis of Quenched Imaging Regimes} \label{sec:quenching_regimes}
\begin{figure}
    \centering
    \includegraphics[width = 0.7\textwidth]{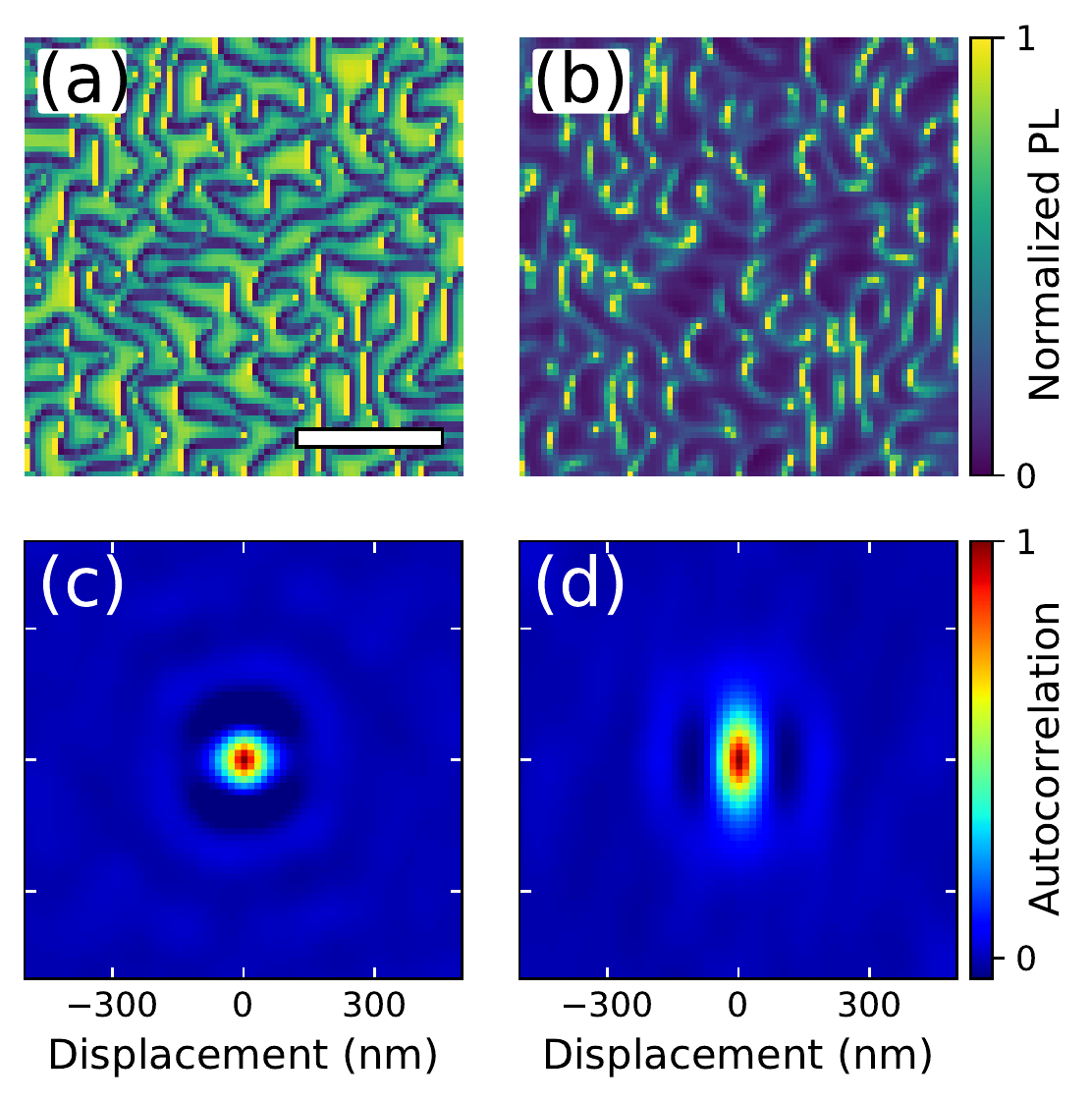}
    \caption{\textbf{Quenched image autocorrelation and directionality.} (a) Quenched images at \numwunit{\dnv{} = 12}{nm} and $I_s = \numwunit{1.6}{mA}$ and (b)  at \numwunit{\dnv{} = 78}{nm} and $I_s = \numwunit{10.5}{mA}$. (Scale bar: $\mathrm{1 \mu m}$) (c, d) Autocorrelation maps of panels a, b, respectively. Quenched maps with low directionality display a a cross-correlation peak with circular simmetry. When the directionality increases, the peak becomes elliptic.}
    \label{fig:Sfig_autocorr}
\end{figure}

\begin{figure}
    \centering
    \includegraphics[width = 0.9\textwidth]{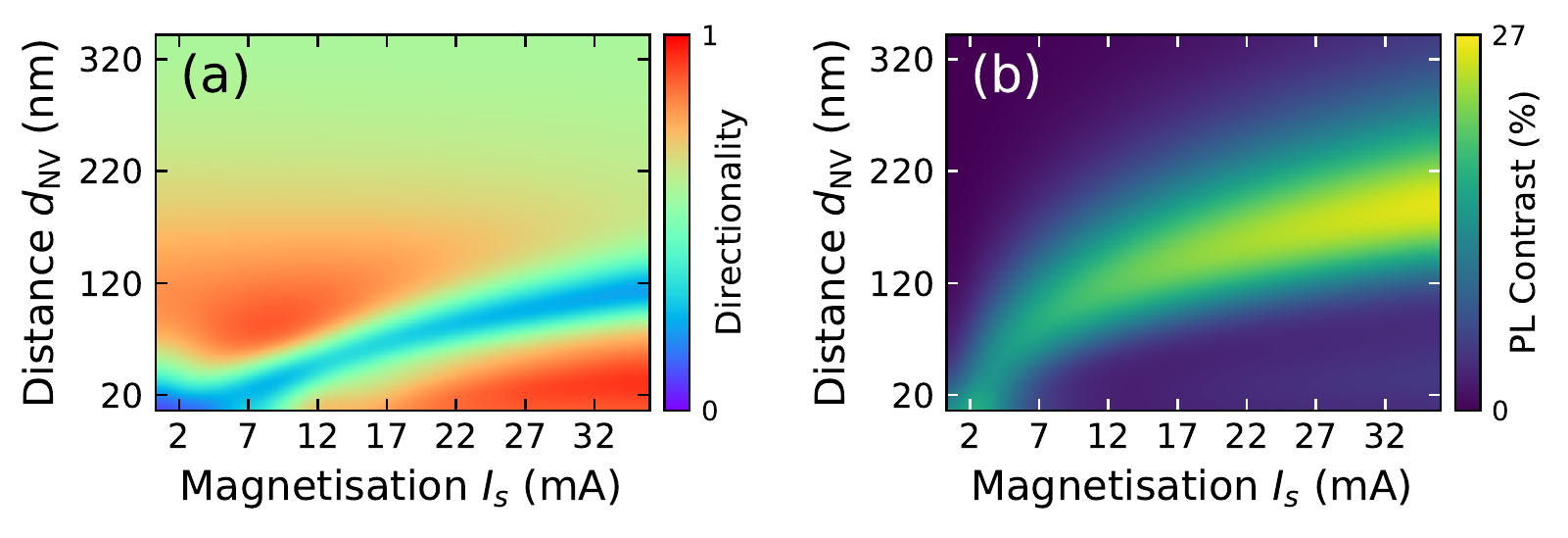}
    \caption{\textbf{Details on Quenched Imaging Regimes}. (a) The directionality of PL features and (b) the PL contrast of a quenched image given as a function of \dnv and $I_s$.
    }
    \label{fig:Sfig_imaging_regimes}
\end{figure}

The diagram in Figure 2 of the main text is constructed based on the directionality of the observed PL features and the contrast of quenched images with different combinations of surface magnetisation $I_s$ and NV-sample distance \dnv. The directionality of the PL features is determined from the auto-correlation of the quenched image (Fig.~\ref{fig:Sfig_autocorr}). The directionality is defined as $1 - r_{min} / r_{max}$, where $r_{\rm min}$ and $r_{\rm max}$ are respectively the minor and the major axis of an elliptical Gaussian fit to the cross-correlation peak. A directionality equal to zero indicates isotropic features  (Fig.~\ref{fig:Sfig_imaging_regimes}(a)), and a value increasing to unity implies increasing anisotropy. The PL contrast is given as $1 - P_{10}/P_{90}$, where $P_{x}$ is the $x^{th}$ percentile of the PL distribution of each quenched image (Fig.~\ref{fig:Sfig_imaging_regimes}(b)).

\begin{figure}
    \centering
    \includegraphics[width = 0.9 \textwidth]{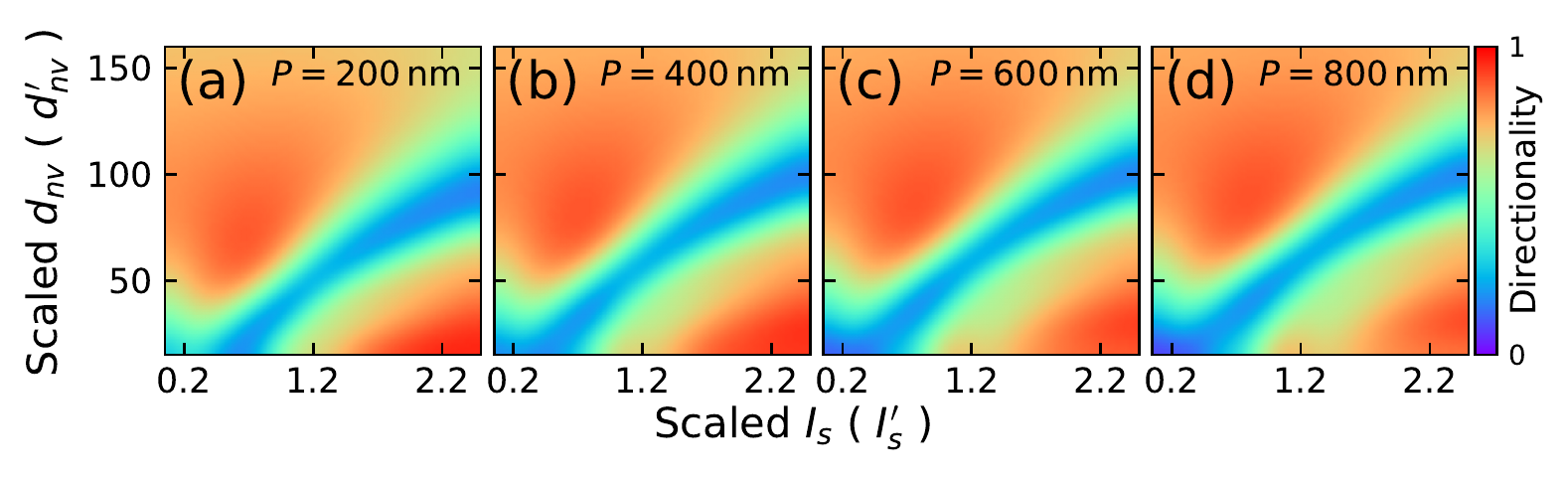}
    \caption{\textbf{Scaled Quenched Imaging Regimes at Varying Domain Periodicity}. The directionality of PL features as a function of scaled $I_s$ ($I_s'$) and scaled \dnv{} ($\dnv{}'$) at domain periodicity, (a) $P=\numwunit{200}{nm}$, (b) $P=\numwunit{400}{nm}$, (c) $P=\numwunit{600}{nm}$ and (d) $P=\numwunit{800}{nm}$. The similar directionality picture indicates that the quench imaging regimes remain valid across different $P$ with appropriate scaling to $I_s$ and \dnv{}.
    }
    \label{fig:SFig_varying_P}
\end{figure}

\begin{figure}
    \centering
    \includegraphics[width = 0.7\textwidth]{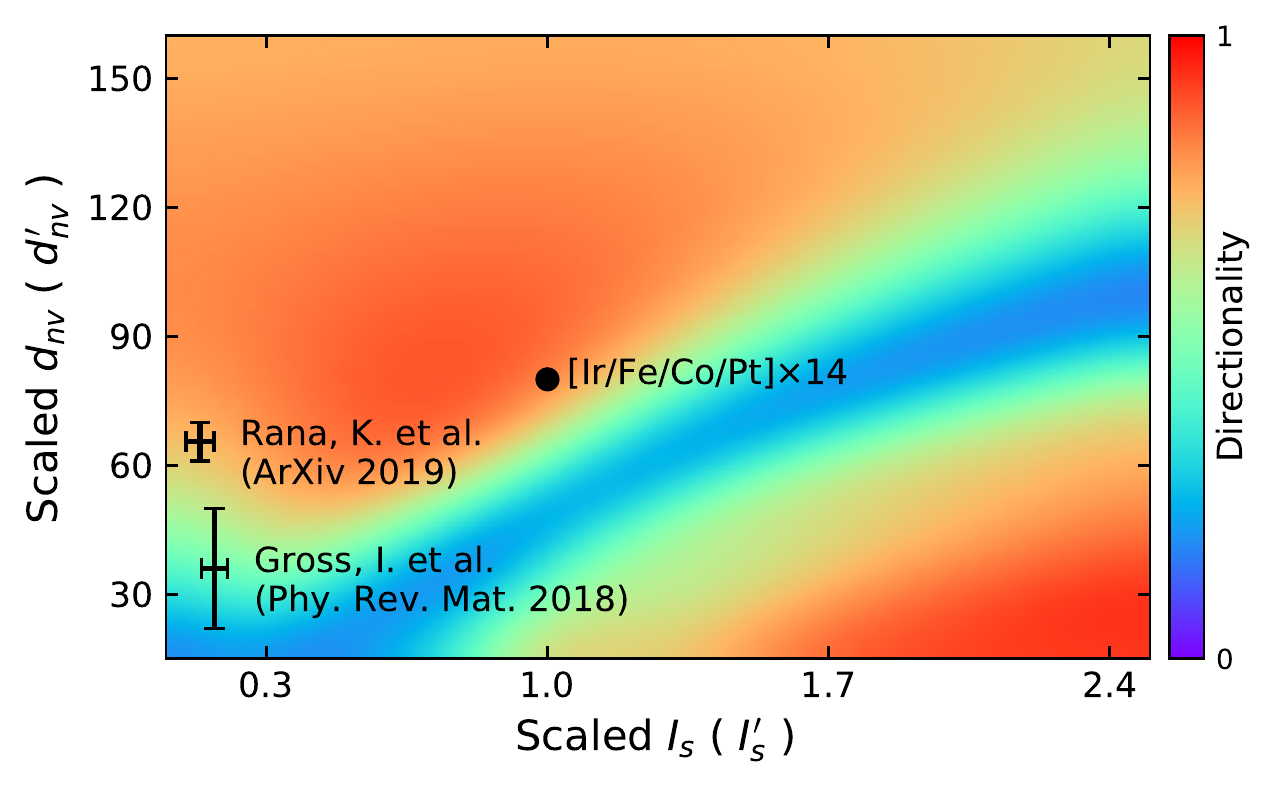}
    \caption{\textbf{Overview of Quenched Imaging on Thin Films}. Previous studies involving quenched imaging of thin films are plotted on the scaled directionality map. The position on the map is based on the \dnv, $I_s$, and $P$ in each study.
    }
    \label{fig:SFig_consolidate}
\end{figure}

Apart from the films magnetisation and \dnv{}, we expect the magnetic field distribution be heavily influenced by the domain periodicity. We show here that the quench imaging regimes put forward in the main paper remain valid at different P, with appropriate scaling of \dnv{} and M. We define the scaled \dnv{} as $\dnv{}' = \dnv{} \times (P/P_0)^{S_d}$, and scaled M as $I_s'= I_s/I_{s,0} \times (P/P_0)^{S_i}$ where $P_0 = \numwunit{407}{nm}$ and $I_{s,0}= \numwunit{12.3}{mA}$ correspond to the value for our sample [Ir($\numwunit{1}{nm}$)/ Co(1)/ Pt(1)]$_{14}$. Scaling factor $S_d$ and $S_i$ are empirically determined to be $-1$ and $-0.8$ $(\sim\sqrt{2/ 3})$. The analytical derivation is however beyond the scope of the paper. The scaled directionality maps at varying $P$ are given in Figure~\ref{fig:SFig_varying_P}. We also include films studied by Gross et al.~\cite{Gross2018} and Rana et al.~\cite{Rana2020} in this framework (Fig.~\ref{fig:SFig_consolidate}). The framework is in good agreement with the work of Gross et al. which observed isotropic PL features. In the study of Rana et al., we are unable to resolve the directionality of the features observed. However, we expect the quenched imaging regime to deviate from our framework as our simulation model does not include exchange bias present in their film. 

\clearpage
\section{NV Sensor Characterisation}  \label{sec:dnv_estimate}
\begin{figure}
    \centering
    \includegraphics[width = .9\textwidth]{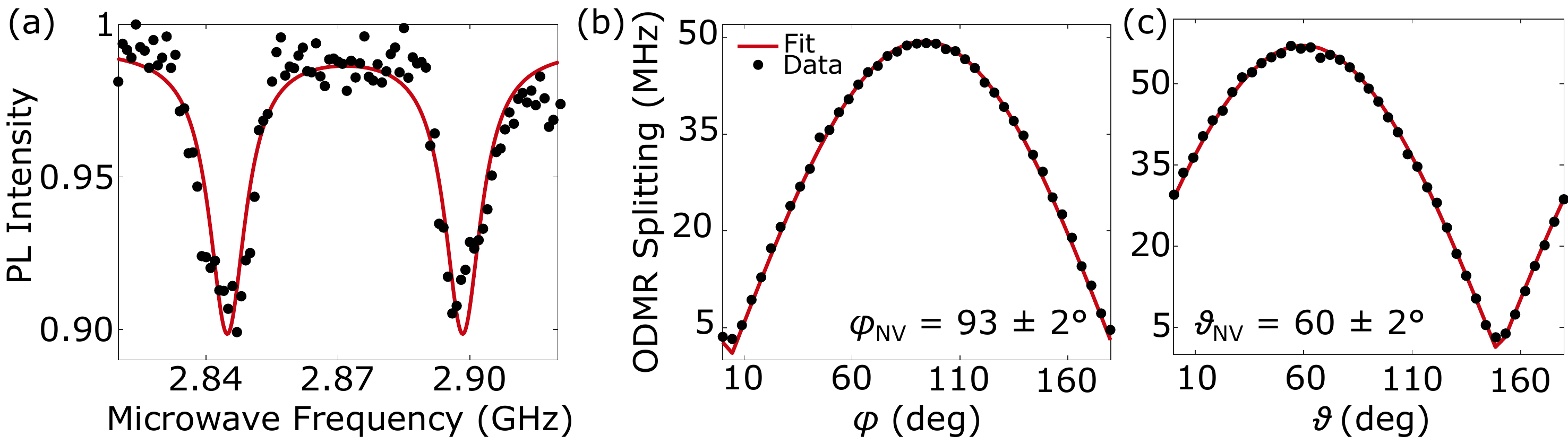}
    \caption{\textbf{Axis measurements of the NV probe.} (a) ODMR spectrum obtained under an external magnetic field. We can observe the splitting of the $\ket{m_s=+1}$ and $\ket{m_s=-1}$ due to the Zeeman effect. We measure a splitting of \numwunit{54}{MHz} which corresponds to a field felt by the NV of about \numwunit{1}{mT}. (b) Measurement of $\phinv$. We fix $\vartheta$ and $\varphi$ is varying. When the ODMR splitting is maximum, $\varphi = \phinv$. (c) Measurement of $\thetanv$. We fix $\varphi$ and $\vartheta$ is varying. $\vartheta = \thetanv$ when the ODMR splitting reaches its maximum value. }
    \label{fig:Sfig_axismeasurement}
\end{figure}

We use a 3-axis Helmholtz coil to apply an external magnetic field $\mathbf{B}$ at varying $\varphi$ and $\vartheta$, with a fixed field strength $\lvert\mathbf{B}\rvert = \numwunit{1}{mT}$. We obtain the ODMR spectra by recording the integrated PL intensity of the NV centre as we sweep the microwave (MW) frequency. In the presence of magnetic field, the ODMR spectrum displays a splitting of the $\ket{m_s=+1}$ and $\ket{m_s=-1}$ states due to the Zeeman effect (Fig.~\ref{fig:Sfig_axismeasurement}(a)). This splitting is proportional to the projection of the magnetic field on the NV axis \unv. The ODMR spectrum is first obtained as a function of $\varphi$ while fixing $\vartheta = \numwunit{90}{^\circ}$ (Fig.~\ref{fig:Sfig_axismeasurement}(b)). The Zeeman splitting is maximum when $\varphi= \phinv$ which in our case is $\phinv = \numwunit{93 \pm 2}{^\circ}$. Next, we vary $\vartheta$ while fixing $\varphi=\phinv$ (Fig.~\ref{fig:Sfig_axismeasurement}(c)). Similarly,  the maximum splitting occurs when $\vartheta= \thetanv$ which we obtain to be $\thetanv= \numwunit{60 \pm 2}{^\circ}$.

\begin{figure}
    \centering
    \includegraphics[width = 0.9\textwidth]{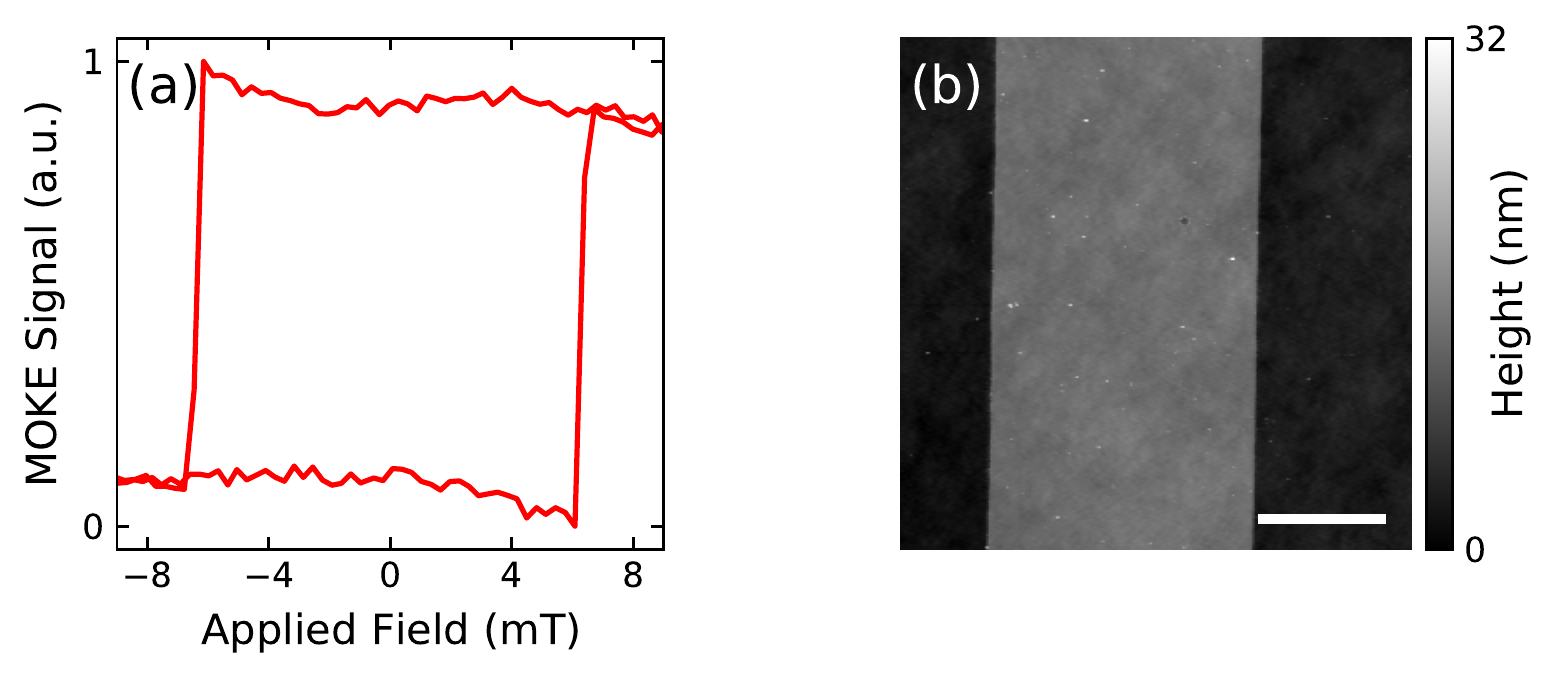}
    \caption{\textbf{Calibration Strip Characterisation.} (a) Intensity of polar MOKE signal of a [Ta/CoFeB/MgO] strip as a function of an out-of-plane magnetic field. (b) Topography image of [Ta/CoFeB/MgO] calibration strip (scale bar: \numwunit{10}{\mu m})}
    \label{fig:Sfig_calibration_sample}
\end{figure}

\begin{figure}
    \centering
    \includegraphics[width = 0.9\textwidth]{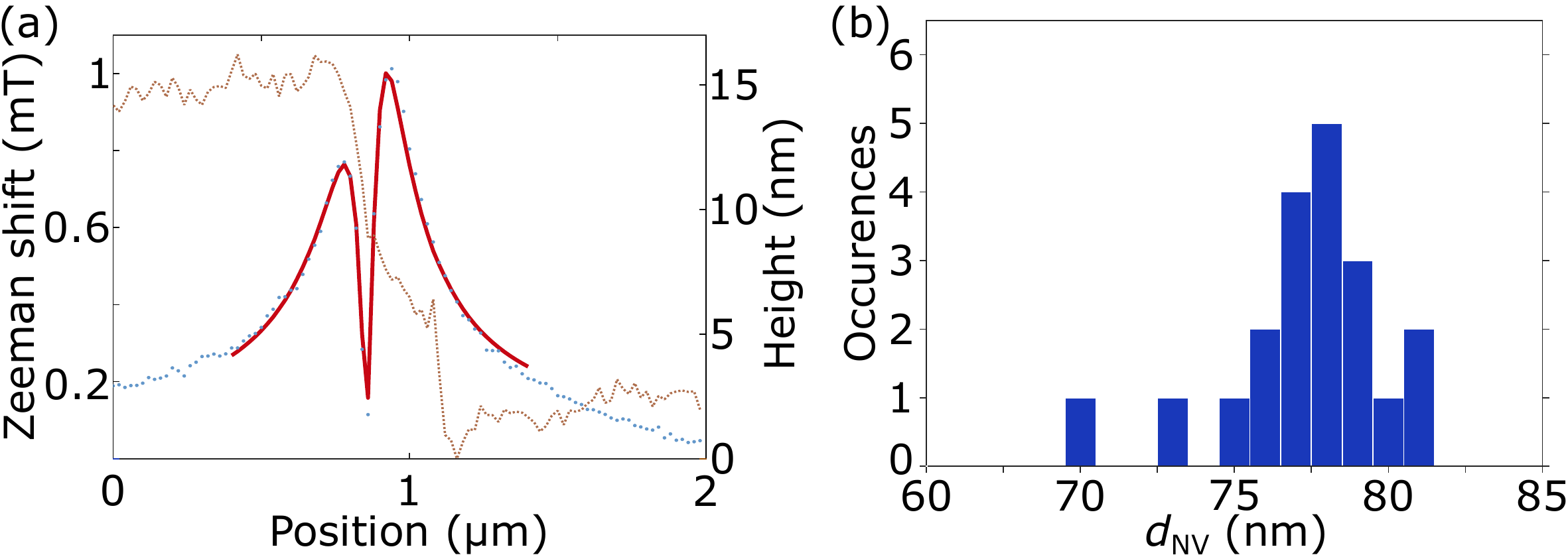}
    \caption{\textbf{Calibration of the NV probe.} (a) We represent on this plot the topography of the edge of a CoFeB magnetic stripe (in brown) and the measured Zeeman shift of the NV ODMR spectrum (in blue) due to the magnetic field emitted at the edge of the stripe. We deduce the value of \dnv{} from the fit (in red) of the Zeeman shift experimentally measured. (b) Histogram distribution of all the NV-sample distances we measured. The average value is $\dnv = \numwunit{77.5}{nm}$ and the standard deviation is $\sigma_{\dnv} \simeq \numwunit{3}{nm}$.}
    \label{fig:Sfig_dNVcalibration}
\end{figure}

We determine the NV-sample distance \dnv{} by measuring with our diamond tip the stray field emitted across the edge of a [Ta/CoFeB/MgO] strip. The out-of-plane magnetic hysteresis is characterised by a MagVision Magneto-Optical Kerr Effect (MOKE) microscope (Vertisis Technology) in the polar sensitivity mode and shows that the magnetisation remains saturated at remanence (Fig.~\ref{fig:Sfig_calibration_sample}). The Zeeman shift of the ODMR spectrum across the edge at remanence is given in Figure~\ref{fig:Sfig_dNVcalibration}(a) (blue dashed curve) and is fitted (red curve) following the procedure devised by Hingant et al.~\cite{Hingant2015} to retrieve the \dnv. We repeat the measurement numerous times along the edge at \numwunit{50}{nm} spacing, and the extracted values are averaged (Fig.~\ref{fig:Sfig_dNVcalibration}(b)). The diamond tip used in this work has a $\dnv = \numwunit{77.5}{nm} \numwunit{\pm 3}{nm}$.

\clearpage
\section{Quenched Imaging with [111] NV Centre} \label{sec:111_nv}
\begin{figure}[b]
    \centering
    \includegraphics[width = \textwidth]{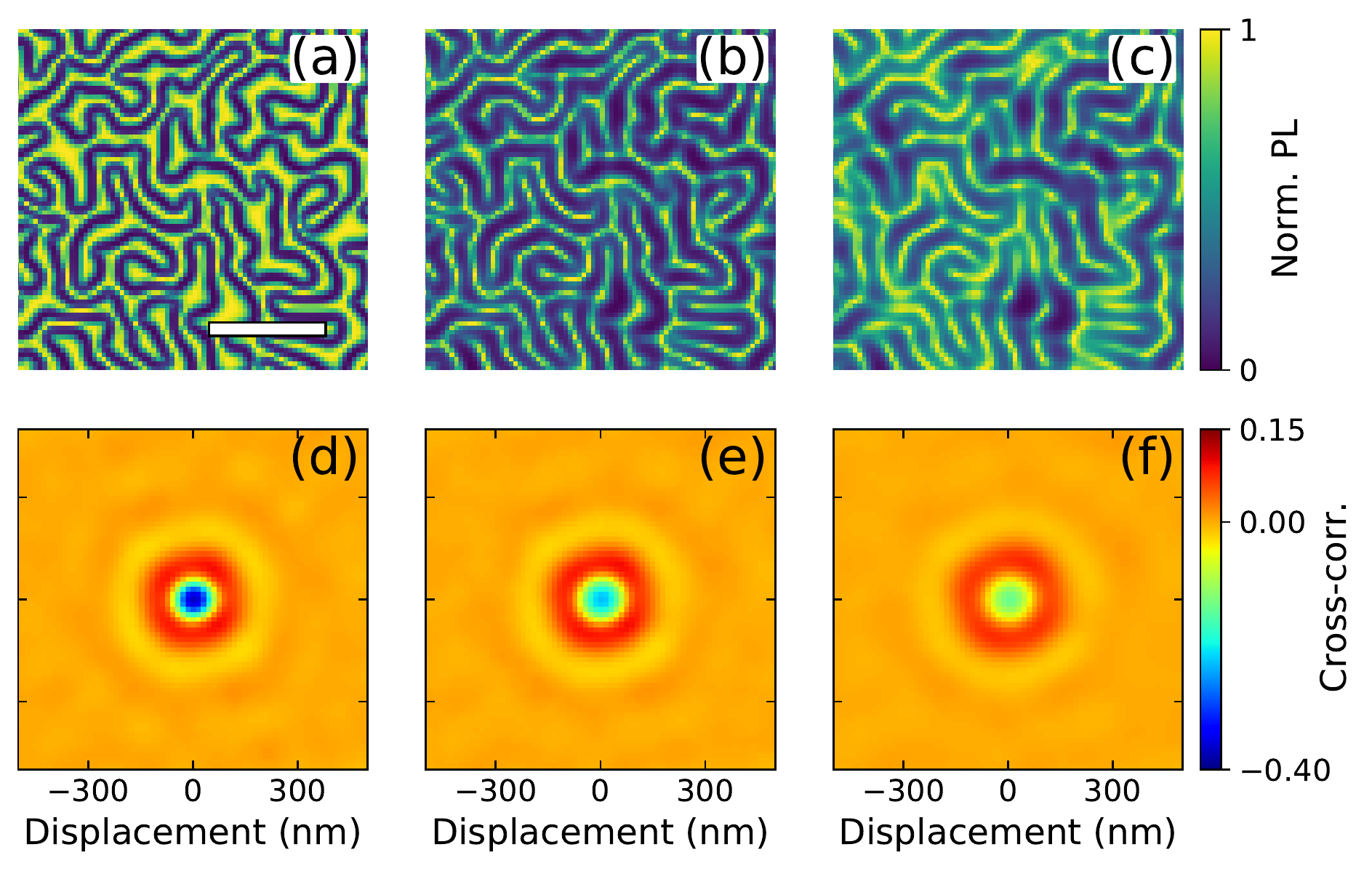}
    \caption{\textbf{Quenching with [111] NV centres}. Quenching maps obtained with NVs with $\thetanv{} = 0^\circ$ on the same area as Figure 2 in the main text  (scale bar: \numwunit{1}{\mu m}). The different maps correspond to (a) $I_s =\numwunit{3.1}{mA}$, $\dnv{} = \numwunit{30}{nm}$, (b) $I_s=\numwunit{9.2}{mA}$, $\dnv{} = \numwunit{84}{nm}$, and (c)  $I_s=\numwunit{15.4}{mA}$, $\dnv{} = \numwunit{162}{nm}$, which correspond to the parameters of Figure 2(b-d) of the main text.
    (d-f) 2D cross-correlation maps of the images in (a-c) with the domain boundaries. The negative correlation at zero displacement indicates a low PL at the boundary. If the displacement increases, the correlation is positive, corresponding to the bright PL observed within the domains.}
    \label{fig:Sfig_diamond111}
\end{figure}
The discussion in the main text focuses on NV centres found in commercially available (100) diamond tips. Quenched maps obtained with NVs with $\thetanv{} = 54.7^\circ$ on samples with out-of-plane magnetic anisotropy give rise to different imaging regimes, as explained in the main text. Notably, there is a range of \dnv{} and $I_s$ where the quenched maps  directionally highlight the domain boundaries. The directionality is due to the non-zero angle between the NV axis and the magnetic anisotropy. Hence, this effect is not present when using NV centres pointing along the [111] axis (i.e. $\thetanv{} = 0^\circ$), hosted in (111)-oriented diamond tips, which have been recently reported~\cite{Rohner2019}. The simulations shown in Figure~\ref{fig:Sfig_diamond111}(a-c), which have been taken at the same \dnv{} and $I_s$ of Figure 2(b-d) of the main text, respectively. At low magnetisation (Fig.~\ref{fig:Sfig_diamond111}(a)), the NV PL is quenched along the domain boundaries (cross-correlation in  Fig.~\ref{fig:Sfig_diamond111}(d)), resulting in a bright image with dark outlines. At higher magnetisation (Fig.~\ref{fig:Sfig_diamond111}(b)) the quenching still traces the domain boundaries (cross-correlation in  Fig.~\ref{fig:Sfig_diamond111}(e)), but also expands further within the domain area. The thin bright lines correspond to the innermost areas of the domains, where the magnetic field is mainly orthogonal to the sample surface and thus aligned with the NV axis. In Figure~\ref{fig:Sfig_diamond111}(c), the combination of large magnetisation and high \dnv{} gives an image similar to Figure~\ref{fig:Sfig_diamond111}(b), but with lower resolution.

\newpage
\section{Domain Coverage Estimation} \label{sec:domain_coverage}
\begin{figure}
    \centering
    \includegraphics[width = \textwidth]{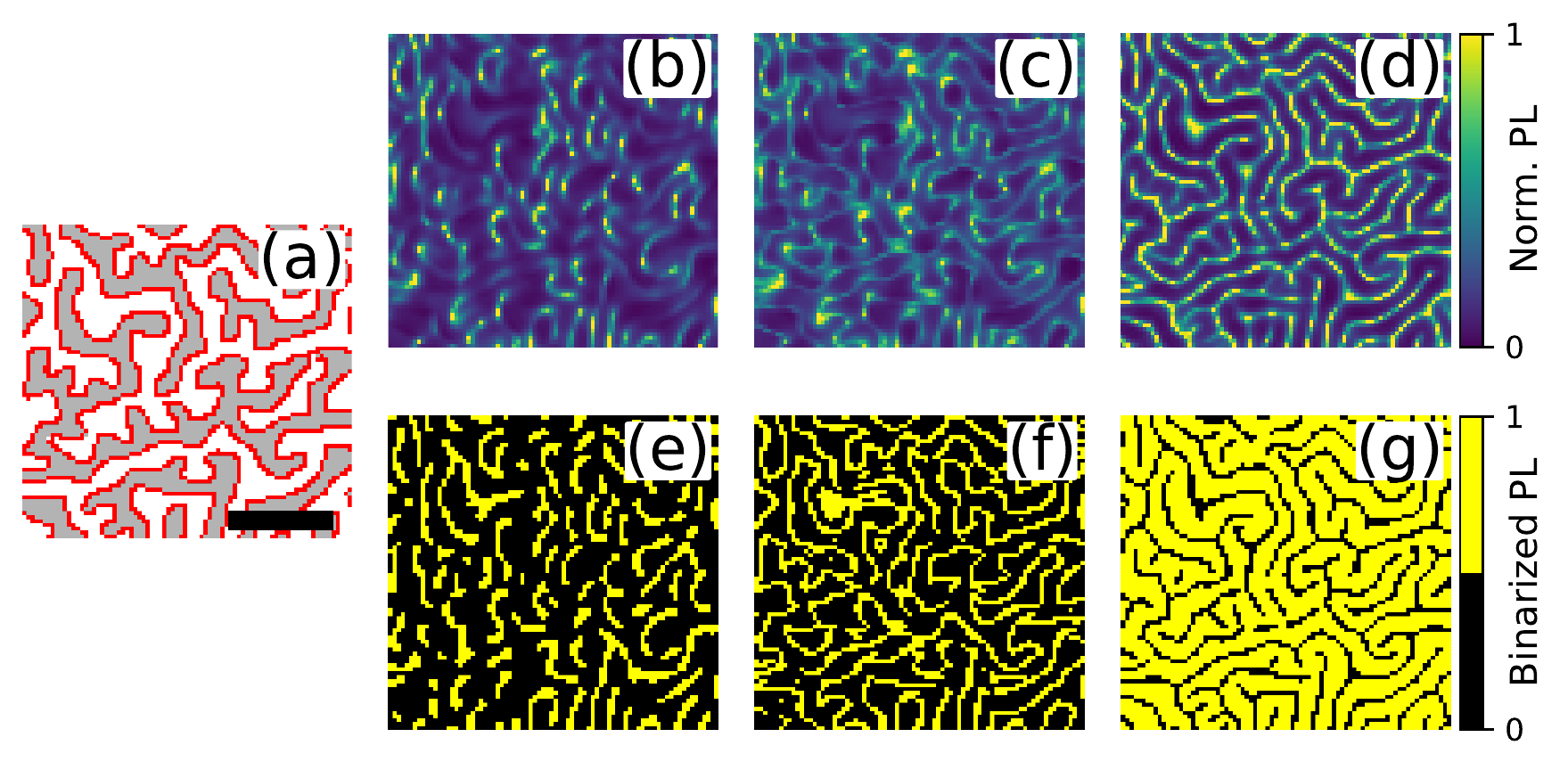}
    \caption{\textbf{Estimation of the domain coverage.} (a) Portion of the binarised MFM scan (background) and domain edges (red pixels) obtained via Canny edge detection (scale bar: \numwunit{1}{\mu m}). Quenched maps of the same area, where (b) is the map taken with an NV with $\thetanv{} = 54.7^\circ$ and $\phinv{} = 0^\circ$, (c) is the reconstructed image with $N=4$ at $\varphi_{max} = 180^\circ$ (see main text), and (d) is acquired with an NV with $\thetanv{} = 0^\circ$. (e-g) are the images obtained by binarising (b-d), respectively.}
    \label{fig:Sfig_bin_reconstruction}
\end{figure}

In order to estimate the percentage of domain boundaries covered by the simulated quenched maps, we first binarize the selected MFM images with Otsu thresholding~\cite{Otsu1979} (a portion is shown Fig.~\ref{fig:Sfig_bin_reconstruction}(a)) and detect the boundaries with the Canny algorithm. The quenched maps are simulated from the stray fields obtained with Mumax3, as explained above. We first simulate the quenched maps with NVs at $\thetanv{} = 54.7^\circ$ and different $\phinv{}$  (Fig.~\ref{fig:Sfig_bin_reconstruction}(b) for $\phinv{} = 0^\circ$). The single images are then registered to the domain boundaries with the ECC image alignment algorithm~\cite{Evangelidis2008}, in order to compensate for the small shift from the domain boundaries induced by the non-zero angle between the magnetic anisotropy and \thetanv{}. The images are then combined as explained in the main text (Fig.~\ref{fig:Sfig_bin_reconstruction}(c) for $N=4$ and $\varphi_{max} = 180^\circ$). Additionally, we simulate a quenched map with an NV at $\thetanv{} = 0^\circ$, which exhibits no shift and no \phinv{}-dependence (Fig.~\ref{fig:Sfig_bin_reconstruction}(d)). The images are then binarized using local gaussian thresholding (Fig.~\ref{fig:Sfig_bin_reconstruction}(e-g)). The binarized images are multiplied to the domain boundaries and integrated to yield the coverage. For experimental quenched maps, the above estimation protocol includes additional thinning and dilation of the binarized images before multiplication. The thinning and dilation process ensures that local deviations between the binarized experimental quenched maps obtained via SNVM and the domain outline retrieved from MFM images are accounted in the coverage estimation. These local deviations are largely due to experimental map nonlinearity, suboptimal image threshold and registration conditions, and MFM perturbation. To estimate the experimental coverage error, we use the 3 smallest structuring element -- a 3 pixel wide diamond, 3 pixel wide square and 5 pixel wide diamond -- for binary dilation. The coverage value is given by the estimation protocol using a 3 pixel wide square dilation structuring element while the coverage bounds are given by the diamond structuring elements.  

\newpage
\section{ Directionality, image reconstruction and boundary coverage } \label{sec:im_reconstruct}
\begin{figure}
    \centering
    \includegraphics[width = \textwidth]{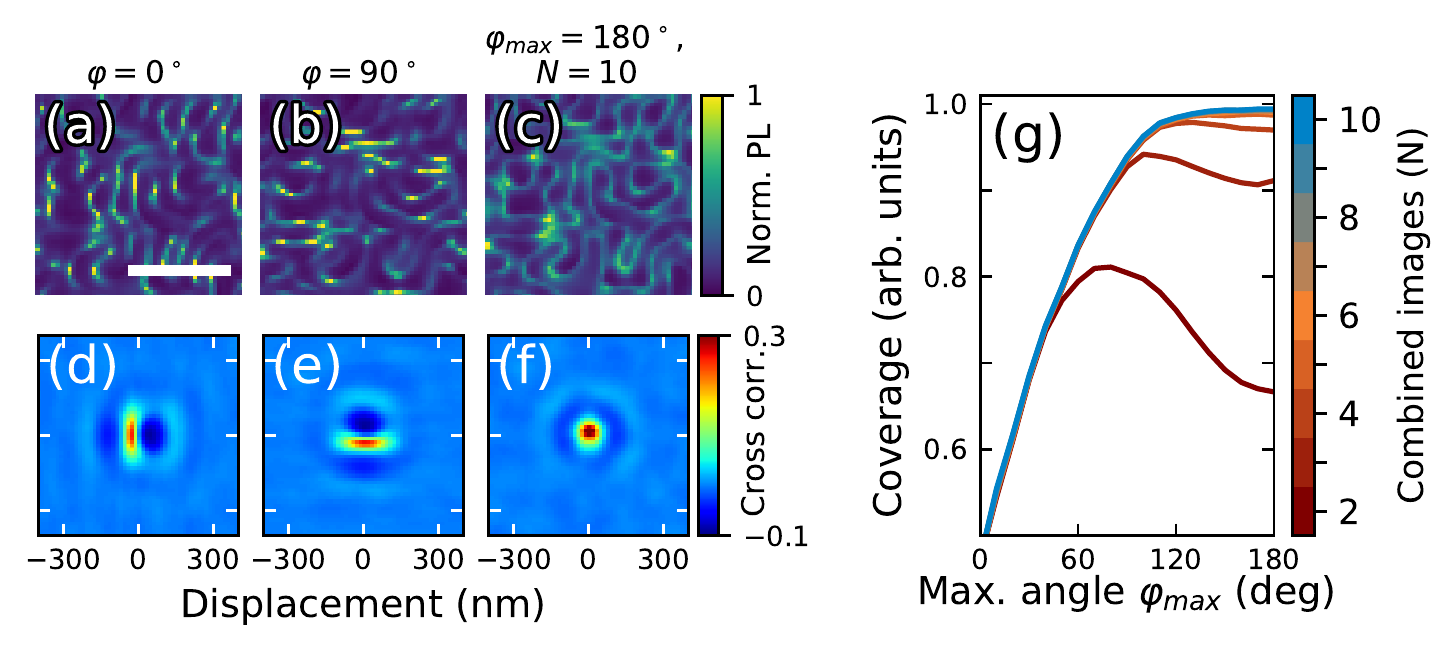}
    \caption{\textbf{Directionality and image reconstruction} (a,b) Simulated quenched maps with $\phinv{} = 0^\circ$ and $\phinv{} = 90^\circ$ and (c) MARe image with $N = 10$ and $\varphi_{max} = 180^\circ$. (Scale bar: \numwunit{1}{\mu m}). (d-f) 2-dimensional cross-correlations between the quenched images in (a-c) and the domain boundaries. For single images (d,e), the cross-correlation shows a positive correlation shifted from the origin in the direction opposite to \unvphi{}, the projection of \unv{} in the sample plane. This indicates that the bright outlines are highly directional and do not occur exactly on top of the domain boundaries. On the contrary, the cross-correlation of the MARe map (f) is isotropic, indicating that most of the boundaries are uniformly covered. (g) Simulations of domain boundary coverage to include $N$ beyond $N=4$.}
    \label{fig:Sfig_crosscorr}
\end{figure}

We can further analyze the properties of quenched maps by studying the 2-dimensional cross-correlation between the maps and the domain boundaries. We do this by simulating the quenched maps starting from the MFM image (Fig.~\ref{fig:Sfig_MFM_threshold}(a-b)), as described above and in the main text. We then calculate the cross-correlation between the maps and the domain boundaries obtained from the MFM map with the Canny edge detection algorithm. The cross-correlation plots (Fig.~\ref{fig:Sfig_crosscorr}(d-e)) show a non-uniform positive correlation peak which is shifted from the origin. The shift is opposite to the direction of the direction of \unvphi{}. This indicates that the directional features on average do not occur on top of the domain boundaries. The shift originates from the non-zero tilt of \unv{} from the normal to the sample plane. This has important consequences for the MARe scheme, since the shift needs to be compensated with image registration algorithms before combining the images (Sec.~\ref{sec:domain_coverage}). In Figure~\ref{fig:Sfig_crosscorr}(c) we show a MARe image with $N = 10$ and $\varphi_{max} = 180^\circ$. The corresponding 2D cross-correlation (Fig.~\ref{fig:Sfig_crosscorr}(f)) displays a circularly symmetric peak, indicating that the reconstructed map is non-directional. 

We also study the option of using more than four images ($N > 4$) to reconstruct the domain morphology. We show the result of the simulations in Figure~\ref{fig:Sfig_crosscorr}(g). As presented in the main text, $N=4$ achieves a peak coverage of about $0.97$ at $\varphi_{max} \approx 120^\circ$. The maximum coverage for $N=5$ is $\approx 0.98$ at $\varphi_{max} \approx 180^\circ$. For $N>5$, the coverage reaches a peak value of $\approx 1$.

\newpage
\section{Imaging with Minimal Perturbation} \label{sec:sample_perturbation}
\begin{figure}
    \centering
    \includegraphics[width = \textwidth]{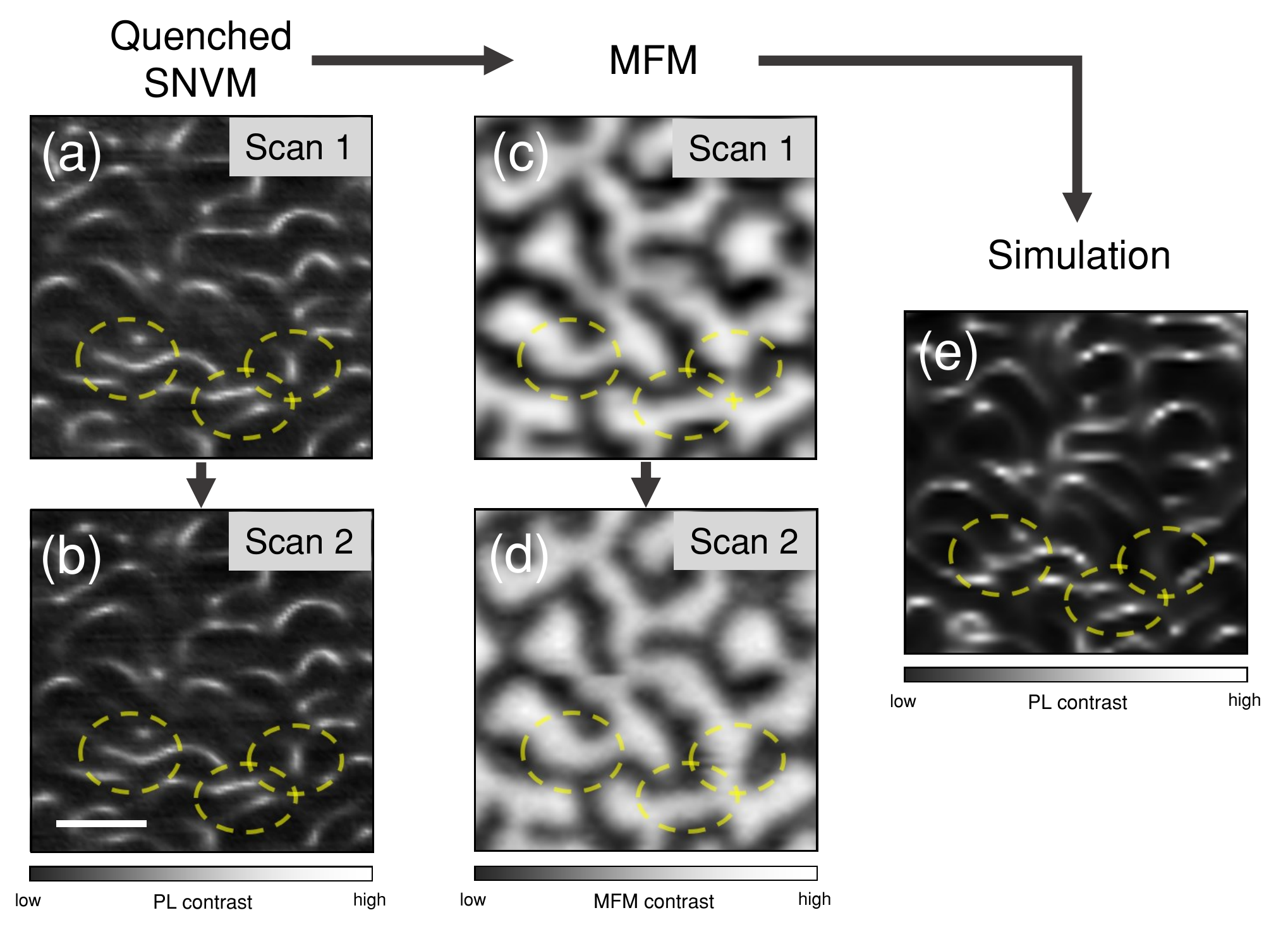}
    \caption{\textbf{Evidence of non-perturbative imaging.} Sequential study of the domain morphology by (a,b) consecutive quenched SNVM imaging, followed by (c,d) consecutive MFM. Areas of perturbation due to consecutive MFM imaging are circled (c, d), while no visible changes are observed in the corresponding areas in the consecutive quenched images (a, b). (e) Simulated quenched image based on second MFM scan 2 (d) shows dissimilar PL features in the circled vicinity as compared to experiments (a, b). (Scale bar: 500 nm)}
    \label{fig:SFig_perturbation_evidence}
\end{figure}

As explained in the main text and in previous studies~\cite{Akhtar2019,Gross2018,Rana2020}, quenched SNVM enables perturbation-free imaging of spin textures. We present here a comparison of repeated quenched SNVM and MFM scans over the same area, which highlight the non-perturbative advantage of quenched SNVM. We first obtain two consecutive quenched images over an area on the sample (Fig.~\ref{fig:SFig_perturbation_evidence}(a,b)), and thereafter another two consecutive MFM images over the same area (Fig.~\ref{fig:SFig_perturbation_evidence}(c, d)). By comparing the quenched and MFM images, we observed areas (circled in Fig.~\ref{fig:SFig_perturbation_evidence}(a-e)) showing non-perturbative consecutive quenched imaging (Fig.~\ref{fig:SFig_perturbation_evidence}(a,b)), but were subsequently perturbed by consecutive MFM scans (Fig.~\ref{fig:SFig_perturbation_evidence}(c,d)). In addition, the quenched image simulated (Fig.~\ref{fig:SFig_perturbation_evidence}(e)) from the MFM image in Figure~\ref{fig:SFig_perturbation_evidence}(d) shows markedly different PL features compared to experiments (Fig.~\ref{fig:SFig_perturbation_evidence}(a,b)) at the vicinity of the highlighted areas (circled in Figure~\ref{fig:SFig_perturbation_evidence}(a), (b) and (e)), reinforcing the non-perturbative advantage of quenched SNVM over conventional MFM. In our case, the MFM probe used for the comparison is a low moment variant from Asylum Research, Oxford Instruments (ASYMFMLM-R2). These observations are however not exhaustive in nature and requires a statistical approach to determine the degree of perturbation induced by MFM over quenched SNVM. A rigorous characterisation is highly non-trivial and involves a vast parameter space including various magnetic material parameters, different laser intensities utilised during quenched SNVM and numerous low-moment probe options for MFM.

\end{document}